\documentclass[a4paper,fleqn]{cas-dc}

\usepackage[numbers,sort&compress]{natbib}
\usepackage{svg}
\usepackage{amsmath, amsfonts}
\usepackage{enumitem}
\usepackage{graphicx}
\usepackage{float}
\usepackage{placeins}
\usepackage{algorithm}
\usepackage[noend]{algpseudocode}
\usepackage{subcaption}
\usepackage{xcolor}
\usepackage{etoolbox}
\usepackage{booktabs}
\usepackage{tabularx}
\usepackage{array}
\usepackage{adjustbox}

\newcolumntype{C}[1]{>{\centering\arraybackslash}p{#1}}

\ExplSyntaxOn
\cs_set:Npn \__make_tbl_caption:nn #1#2
{
	\l_tbl_align_tl
	\skip_vertical:N \l_tbl_abovecap_skip
	{\parbox{\dimexpr(\l_tbl_width_dim)}
		{\rightskip=0pt\sffamily\small\textbf{\color{scolor}#1:~ }#2\par\vskip4pt }}
	\skip_vertical:N \l_tbl_belowcap_skip
}

\cs_if_exist:NF \g_stm_orcid_seq
{ \seq_new:N \g_stm_orcid_seq }

\cs_if_exist:NF \l_stm_au_orcid_tl
{ \tl_new:N \l_stm_au_orcid_tl }

\cs_if_exist:NF \orcidauthor
{ \NewDocumentCommand{\orcidauthor}{mm}{} }
\ExplSyntaxOff

\def\tsc#1{\csdef{#1}{\textsc{\lowercase{#1}}\xspace}}
\tsc{WGM}
\tsc{QE}
\tsc{EP}
\tsc{PMS}
\tsc{BEC}
\tsc{DE}

\begin{document}
	
	\let\WriteBookmarks\relax
	\shortauthors{}
	
	\hypersetup{
		colorlinks=true,
		linkcolor=blue,
		citecolor=blue,
		urlcolor=blue
	}
	
	\makeatletter
	\renewcommand{\@biblabel}[1]{\textcolor{black}{[#1]}}
	
	\patchcmd{\@bibitem}
	{\ignorespaces}
	{\color{blue}\ignorespaces}
	{}{}
	
	\patchcmd{\@lbibitem}
	{\ignorespaces}
	{\color{blue}\ignorespaces}
	{}{}
	\makeatother
	
	\title [mode = title]{NeuroStrata: An Electroencephalographic Connectivity-Aware Deep Representation Learning Framework for Dynamic Brain Network Analysis of Mental Stress}
	\shorttitle{Engineering Applications of Artificial Intelligence}
	\shortauthors{S. Acharya et al.}
	
	\author[1]{Sayantan Acharya}
	\author[1]{Hamzeh Asgharnezhad}
	\author[1]{Abbas Khosravi}
	\author[1]{Douglas Creighton}
	\author[1]{Roohallah Alizadehsani\textsuperscript{*}}
	\author[2,3]{U Rajendra Acharya}
	\address[1]{Institute for Intelligent Systems Research and Innovation (IISRI), Deakin University, Waurn Ponds, VIC, Australia.}
	\address[2]{Centre for Health Research, University of Southern Queensland, Springfield, Australia.}
	\address[3]{School of Science, Engineering and Digital Technologies, University of Southern Queensland, Springfield, Australia.}
	
	
	\begin{abstract}
		This study introduces NeuroStrata, a connectivity-aware deep representation learning framework for electroencephalogram (EEG)-based mental stress analysis using Time-Varying Partial Directed Coherence (TV-PDC). Unlike conventional EEG classification pipelines that rely on static features, the proposed framework models the temporal evolution of frequency-specific directed connectivity across distributed brain regions. EEG signals from the 32-channel SAM 40 dataset, recorded during mental arithmetic tasks, were used to produce frequency-specific TV-PDC connectivity maps. These maps were processed using pretrained Convolutional Neural Networks (CNNs) and Vision Transformers (ViTs) to extract deep connectivity embeddings, which were subsequently classified using lightweight machine learning (ML) models. Experimental results show that beta-connectivity exhibits the highest discriminative capability, achieving a peak accuracy of 97.3\% using the LAION-CLIP-ViT-L14 backbone with a Support Vector Machine (SVM) classifier, while alpha-connectivity provides consistently stable performance across model configurations. Connectivity feature importance analysis reveals prominent frontal-driven alpha influences and centrally integrated beta connectivity patterns, reflecting the engagement of regulatory and sensorimotor networks during stress. Temporal evaluation further indicates that classification performance stabilizes in mid-to-late temporal windows, suggesting the progressive consolidation of stress-related connectivity signatures. From an artificial intelligence perspective, the framework implements a connectivity-aware deep representation learning strategy that integrates time-varying effective connectivity modelling with deep embedding extraction using pretrained CNN and ViT architectures to encode dynamic directed brain network structure. From an engineering application perspective, it enables structured evaluation of stress-modulated connectivity dynamics through interpretable directed network analysis, temporal window stability characterization, and a unified multi-stage EEG connectivity processing pipeline for automated inference of task-evoked cognitive stress states.
		
	\end{abstract}
	
	\begin{keywords}
		Mental stress \sep Electroencephalogram \sep Effective connectivity \sep Partial directed coherence \sep Deep learning \sep Machine learning \sep Multi-level classification
	\end{keywords}
	
	\maketitle
	
	\section{Introduction}
	
	Mental stress is a progressive ‘fight or flight’ response that is mediated by cortical activities but involves numerous psycho-physiological (psy-phy) processes when individuals experience compulsion or agitation during cognitively demanding circumstances \citep{arsalan2022human}. From a cortical perspective, this response involves dysregulation across multiple neural systems, including cortical–limbic interactions and autonomic control pathways \citep{arsalan2022human, kivimaki2018effects}. Situations that elicit stress may be physical, arising from illness or acute injury; social, driven by financial hardship or adverse living conditions; emotional, associated with attributes such as job loss or bereavement; or traumatic, resulting from persistent or distressing past events. Sustained activation of the neural stress pathways leads to chronic stress, which induces long-term neurophysiological alterations that imply cardiovascular dysfunction \citep{kivimaki2018effects}, sleep deprivation \citep{nollet2020sleep}, Post Traumatic Stress Disorder (PTSD) \citep{shalev2017post} and substance abuse \citep{sinha2024stress}. Therefore, mental stress has become a pervasive condition with significant psychological and physiological health complications, thereby highlighting the critical requirement for its methodical understanding and quantification.
	
	Previously, mental stress has been assessed using self-reported questionnaire-based approaches. Among the questionnaires, the Perceived Stress Scale (PSS) is considered one of the most reliable tools for assessing perceived stress \citep{harris2023perceived}. However, stress perception is inherently subjective and varies considerably across individuals. This individual variability poses a substantial challenge for accurate stress-level classification, as measurement accuracy depends critically on the assessment methodology. Researchers have evaluated mental stress by using two principal approaches: psychological self-reported questionnaires and objective physiological markers.
	
	Nevertheless, psychological questionnaire-based assessment is insufficient for a comprehensive evaluation of stress, as stress constitutes a dynamic fight-or-flight response that originates in the brain and engages multiple interacting physiological processes. Self-reported assessment is inherently subjective and influenced by individual perception, emotional awareness, and situational context, leading to variability and inaccurate measurement. Since the brain coordinates the stress response through distributed cortical and subcortical neural networks, exclusive reliance on self-reported questionnaires might risk overlooking the underlying neural dynamics. Moreover, psychological responses often vary over time and may fail to capture the rapid physiological alterations that occur during stress exposure. Consequently, contemporary research integrates subjective assessments with objective physiological markers, including Electroencephalography (EEG) \citep{al2017stress}, heart rate variability \citep{ahn2019novel}, skin conductance \citep{affanni2020wireless}, salivary cortisol levels \citep{ali2020salivary}, and blood pressure \citep{fischer2017blood}. This multi-modal approach, combining neural, physiological, and psychological measures, is crucial for accurately understanding the complexity of human stress responses and their variation across individuals.
	
	\begin{figure*}[t]
		\centering
		\includegraphics[width=\columnwidth]{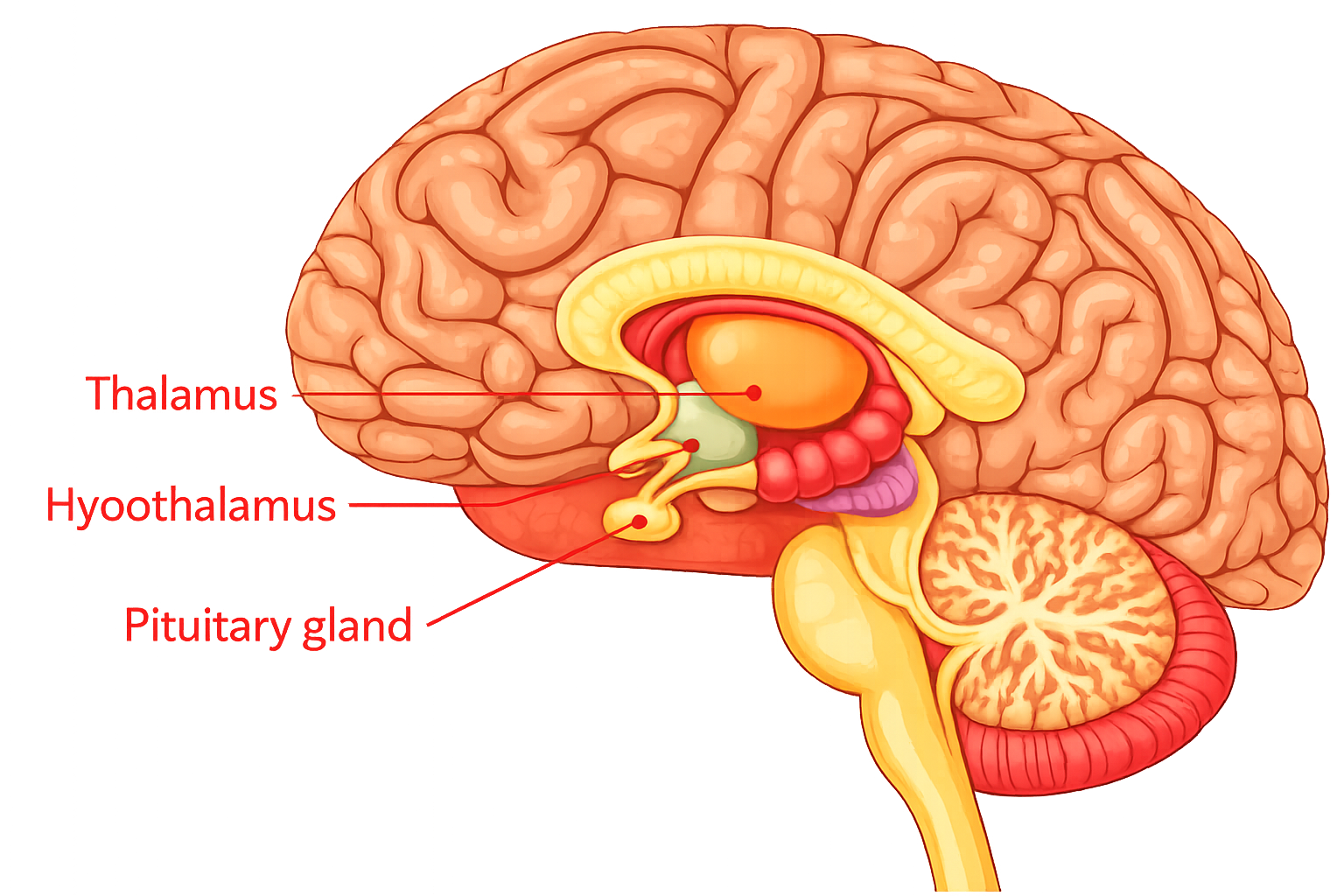}
		\caption{Key brain regions involved in stress processing.}
		\label{fig:brain_regions}
	\end{figure*}
	
	Researchers have demonstrated that mental stress is centrally mediated by the Sympathetic Nervous System (SNS) and regulated through key cerebral structures, including the hypothalamus, pituitary gland, and adrenal cortex, which constitutes the Hypothalamic–Pituitary–Adrenal (HPA) axis \citep{hu2015signal}. As shown in \autoref{fig:brain_regions}, the highlighted brain regions demonstrate the coordinated interaction between thalamic sensory processing and hypothalamic–pituitary neuro-endocrine regulation, which can alter stress-related neural dynamics and inter-connectivity patterns \citep{keshmiri2021conditional, alyan2021frontal, chae2021relationship, al2021cognitive}.

	Consequently, EEG is employed in this study to investigate the stress-induced dynamic alterations in cognitive processing. EEG is an effective modality for stress assessment due to its high temporal resolution, its ability to support frequency-specific brain analysis, and its capability to capture moment-to-moment changes in functional activity and connectivity across distributed brain regions \citep{al2017stress, darzi2022brain, alonso2015stress}. A complex neural signal, such as EEG, can be acquired non-invasively using scalp electrodes, offering a cost-effective and user-friendly recording setup. In this study, the EEG frequency spectrum is partitioned into five distinct bands, as summarized in \autoref{tab:eeg_spectrum}. Each EEG band is associated with specific mental states \citep{al2015mental}. Increased beta power is commonly correlated to heightened alertness and arousal, whereas higher alpha activity reflects relaxed states, and theta oscillations are typically observed during sleep or drowsiness \citep{wang2014emotional}. In the context of stress-inducing cognitive tasks, alpha and beta bands play a critical role, reflecting mechanisms associated with attentional regulation, cognitive processing, and increased vigilance \citep{al2015mental, wang2014emotional}. Earlier investigations of mental stress predominantly relied on Power Spectral Density (PSD)–based features extracted from frontal EEG channels \citep{darzi2022brain, ameera2019analysis}. To date, the highest stress classification accuracies of 99.94\%, 99.93\%, and 99.75\% have been reported using Linear Discriminant Analysis (LDA), k-Nearest Neighbor (kNN), and cubic Support Vector Machine (SVM) classifiers respectively, by employing PSD-driven metrics such as median frequency, spectral moments, Root Mean Square (RMS), and modified frequency mean under relaxed and stressed conditions \citep{attallah2020effective}. Nevertheless, the stress development process is intrinsically linked to coordinated activity across multiple brain regions, as the brain is responsible for perceiving threats or challenges and initiating the stress response \citep{darzi2022brain, ameera2019analysis, khosrowabadi2018stress}. Consequently, brain connectivity analysis offers a promising avenue for capturing functional and effective neural interactions, thereby enabling a more accurate representation of brain dynamics by elucidating how brain regions communicate and influence one another during the stress-building process \citep{van2019cross}.
	
	Recent research has increasingly focused on Deep Learning (DL)-oriented frameworks for EEG-based mental stress detection, with particular emphasis on end-to-end attribute learning and automated classification. Hybrid network architectures integrating convolutional and recurrent components have been widely investigated, wherein convolutional Neural Networks (CNNs) facilitate hierarchical spatial feature extraction, while Long Short-Term Memory (LSTM) or Bidirectional LSTM (BLSTM) layers capture temporal dependencies inherent in EEG time-series signals \citep{malviya2022novel}. These approaches have demonstrated promising classification performance in both binary and multi-class stress recognition scenarios \citep{malviya2022novel}. Furthermore, adaptive dynamic ensemble learning strategies incorporating Correlation-based Importance Score (CIS) feature selection have been proposed to optimally select discriminative EEG features and classifiers, yielding a competitive and subject-consistent performance under varying experimental conditions \citep{malviya2023cis}. Concurrently, RNN–based pipelines, particularly LSTM-based architectures, have been validated on benchmark datasets such as WESAD, achieving strong stress-level detection performance using multi-modal physiological and EEG-derived features \citep{malviya2023mental}. Despite their high predictive capability, these DL-centric models primarily prioritize classification accuracy and remain largely model-driven, offering limited interpretability regarding stress-modulated, frequency-specificity, and directionally dynamic brain network interactions. This limitation underscores the necessity of connectivity-aware, neurophysiologically interpretable stress analytical frameworks.
	
	Previously, researchers have employed various brain connectivity metrics, such as Magnitude Square Coherence (MSC), Coherence, Phase–slope Index (PSI), Canonical Correlation Analysis (CCA), and Mutual Information (MI), to classify stress. MSC-based features have demonstrated higher classification accuracy when combined with SVM, while outperforming PSI, CCA, and PSD-based metrics \citep{darzi2022brain}. Nevertheless, a key limitation of MSC and conventional coherence-based analysis is their heightened sensitivity to fluctuations in signal power and phase coupling \citep{subhani2016difference, balconi2018functional}. Such sensitivity can lead to the detection of apparent connectivity even in the absence of genuine functional interactions between the signals, thereby increasing the risk of identifying spurious connections while masking true connectivity patterns \citep{rho2022valence}. In contrast, CCA is effective in assessing the impact of psychological stress through cross-covariance-based evaluation \citep{al2017assessment}. However, its applicability is limited to linear relationships, which may fail to capture the complex nonlinear interactions present in the signals \citep{al2017assessment}. Previous studies have used PSI and generalized Partial Directed Coherence (gPDC) for stress classification, where PSI exhibited lower classification accuracy than PDC and Directed Transfer Function (DTF) \citep{khosrowabadi2018stress}. This limitation arises because PSI primarily quantifies phase synchronization and does not explicitly identify directed information flow among brain regions \citep{bastos2016tutorial}. Moreover, existing PDC-based studies typically rely on static connectivity embeddings that do not capture directed influences among EEG channels across temporal windows, potentially overlooking dynamic changes in brain networks. Simultaneously, Phase Locking Value (PLV) has been employed to estimate functional connectivity features; however, it yields static measures of phase synchronization between brain regions and does not capture temporal dynamics \citep{hag2021eeg}. Similarly, Amplitude Envelope Correlation (AEC) has been used as an alternative functional connectivity metric; however, it does not capture temporal-window variability or directional influences among brain regions \citep{vanhollebeke2023effects}. 
	
	These limitations highlight the requirement for a structured, methodical framework that can capture directed, frequency-specific, and multi-level neural interactions while supporting robust learning-based stress inference. To address these challenges, ‘NeuroStrata’ is introduced in this research. NeuroStrata is a connectivity-aware deep representation learning framework for EEG-based stress analysis. The primary objective of this work is to develop a layered analytical pipeline that integrates time-varying effective connectivity estimation with deep neural representations and lightweight machine learning classifiers to enable hybrid modelling of mental stress dynamics. Specifically, the framework combines Time-Varying Partial Directed Coherence (TV-PDC) connectivity estimation with pretrained deep embedding models to learn discriminative representations from dynamic directed connectivity patterns. By transforming time-varying connectivity into dynamic connectivity maps, NeuroStrata facilitates deep representation learning of effective brain networks while preserving interpretability through connectivity-based analysis. In summary, the contributions of this paper are outlined as follows:
	\begin{itemize}[leftmargin=*, itemsep=2pt]
		\item Connectivity-aware stress analysis framework:
		The proposed NeuroStrata pipeline integrates EEG time-varying effective connectivity with deep representation learning, enabling the extraction of informative embeddings from directed EEG connectivity patterns for mental stress analysis.
		\item Deep representation learning from dynamic connectivity maps:
		Pretrained DL models, including Convolutional Neural Networks (CNNs) and Vision Transformers (ViTs), are employed to learn discriminative representations from TV-PDC connectivity maps, which are subsequently classified using lightweight Machine Learning (ML) models for 3-level mental stress recognition.
		\item Interpretability through connectivity feature analysis:
		A comprehensive evaluation of top-ranked TV-PDC connections is conducted to identify the most influential stress-modulated information flow pathways and to characterise the dynamic reconfiguration of directed brain connectivity across cortical regions.
		\item Temporal dynamics of stress-related connectivity:
		The impact of temporal segmentation is investigated to understand how evolving connectivity patterns across successive time windows (T1-T7) influence DL-ML classification performance, providing insights into the temporal stability and discriminative power of stress-based neural interactions.
	\end{itemize}
	
	This paper is structured as follows: the Materials and Methods section describes the EEG dataset, stress task, and data labelling approach, followed by the NeuroStrata pipeline, which includes the computation of the TV-PDC algorithm. The Classification Results section presents the results obtained using various DL–ML pipelines, followed by the evaluation of TV-PDC connectivity patterns, the impact of temporal windows and the Discussion. The paper concludes with the Conclusion section.
	
	\section{Materials and methods}
	\label{2: materials}
	
	EEG recordings from the SAM 40 dataset were utilized in this study to investigate time-varying brain connectivity under stress conditions \citep{ghosh2022sam}. The dataset comprises multichannel EEG signals acquired during task execution, along with corresponding participant responses. Mental Arithmetic Task (MAT) based EEG trials were categorized into three stress levels: Relaxed, Low Stress, and High Stress. The EEG bands defined in the SAM 40 dataset are summarized in \autoref{tab:eeg_spectrum}.
	
	
	\begin{table}[!hp]
		\centering
		\renewcommand{\arraystretch}{1.2}
		\caption{Details of EEG spectrum used in this work.}
		\label{tab:eeg_spectrum}
		\begin{tabular}{ccccc}
			\toprule
			\textbf{Delta} & \textbf{Theta} & \textbf{Alpha} & \textbf{Beta} & \textbf{Gamma} \\
			\midrule
			0.5--4 Hz & 4--8 Hz & 8--13 Hz & 13--30 Hz & 30--45 Hz \\
			\bottomrule
		\end{tabular}
	\end{table}
	
	\begin{figure*}[t]
		\centering
		\includegraphics[width=0.5\textwidth]{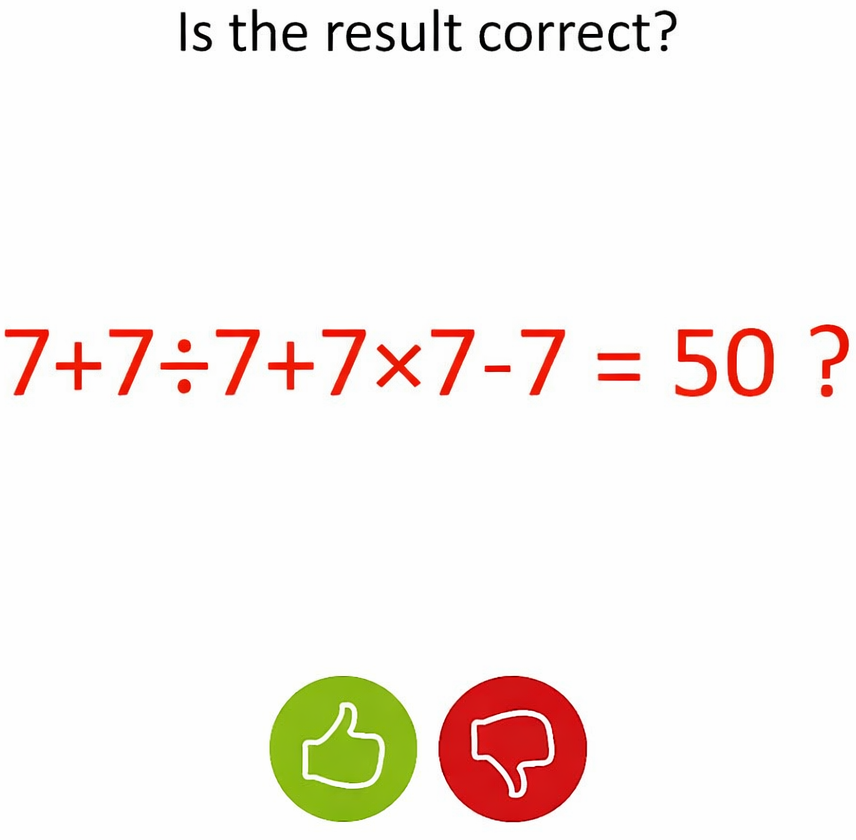}
		\caption{The MAT paradigm for stress task \citep{ghosh2022sam}.}
		\label{fig:MAT_task}
	\end{figure*}
	
	\subsection{EEG dataset}
	
	The SAM 40 dataset consists of EEG recordings from 40 participants (26 males and 14 females; mean age: 21.5 years), sampled at 128 Hz \citep{ghosh2022sam}. The EEG frequency bands employed in this study are consistent with the defined bands in the SAM 40 dataset, as summarized in  \autoref{tab:eeg_spectrum}. The EEG data were collected while participants performed three stress-inducing tasks: Symmetric Mirror Image (SMI) recognition, the Stroop Color–Word Task (SCWT), and MAT. The EEG recordings were acquired using a 32-channel Emotiv EPOC Flex gel kit configured with CMS/DRL reference scheme \citep{ghosh2022sam}. No further re-referencing was performed beyond the dataset's prescribed reference protocol. Before the analysis, the dataset was preprocessed to remove baseline drifts by subtracting the estimated average trend using the Savitzky–Golay filtering technique \citep{ghosh2022sam}. Additionally, physiological artifacts were attenuated through wavelet thresholding \citep{ghosh2022sam}. 
	
	\subsection{Stress task}
	
	For time-varying connectivity analysis, the MAT-based EEG recordings from the SAM 40 dataset were considered \citep{ghosh2022sam}. The MAT is a well-established paradigm in psy-phy and is widely known for reliably eliciting stress-related neural changes, particularly in brain regions associated with executive function, attention, and cognitive control \citep{katmah2021review}. During each trial, participants were required to mentally solve arithmetic problems and indicate whether the displayed solution was correct using a thumbs-up or thumbs-down response \citep{ghosh2022sam}. As illustrated in \autoref{fig:MAT_task}, each trial consisted of six arithmetic problems involving different operators, yielding 25 seconds of EEG data per trial. The selection of the MAT was further motivated by the more consistent EEG alterations observed across participants during this task, whereas the SCWT and SMI tasks have been reported to produce comparatively less consistent EEG responses in the SAM 40 dataset \citep{ghosh2022sam}.
	
	\subsection{Data labeling}
	
	The MAT-based EEG trials were labeled as Low Stress and High Stress based on the SAM 40 dataset’s self-reported stress ratings on a 10-point Likert scale. For each participant, three MAT trials were recorded, with the trial receiving the highest stress rating (e.g., 7/10) labeled as High Stress and the lowest-rated trial (e.g., 2/10) labeled as Low Stress. The dataset also includes three separate Relaxed-state EEG recordings acquired before each MAT session for all participants \citep{ghosh2022sam}. To ensure reliable labeling, EEG data from five participants were excluded because their stress ratings exhibited insufficient variability to distinguish Low Stress from High Stress, resulting in a final dataset comprising 35 participants for analysis. 
	
	\section{NeuroStrata}
	\label{3:neurostrata}
	
\begin{figure*}[t]
		\centering
		\includegraphics[width=0.98\textwidth]{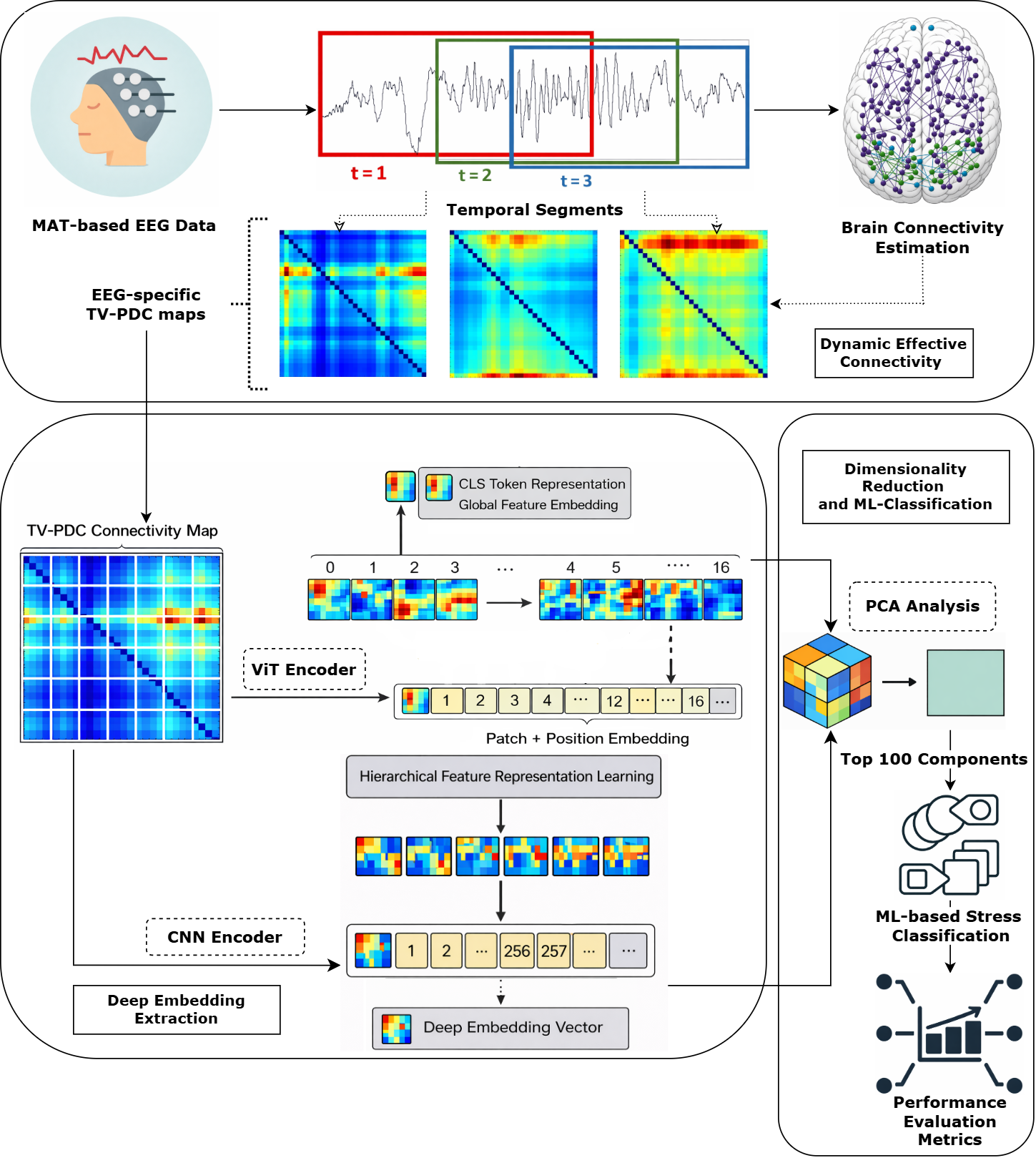}
		\caption{Proposed NeuroStrata framework.}
		\label{fig:TL_pipeline}
	\end{figure*}

	In this study, NeuroStrata is introduced and operationalized as a multi-stage analytical pipeline for extracting and modeling time-varying effective connectivity patterns from multichannel EEG data. As shown in \autoref{fig:TL_pipeline}, the pipeline begins with short-time window segmentation of pre-processed MAT-based EEG signals from the SAM 40 dataset, followed by estimation of frequency-specific directed information flow maps by computing the Time-Varying Partial Directed Coherence (TV-PDC) protocol. TV-PDC maps capture causal information flow among EEG channels across successive temporal windows. Each TV-PDC map is represented as a directed connectivity matrix corresponding to a specific frequency band and time segment, thereby encoding brain dynamics associated with the processing of mental stress. These TV-PDC connectivity maps are subsequently processed by pretrained CNNs and ViTs, employed as deep feature extractors to obtain compact, discriminative features that capture both localized connectivity patterns and global brain network dependencies. In \autoref{fig:TL_pipeline}, the ViT encoder captures global dependencies in TV-PDC patterns via self-attention, producing a Classification (CLS) token representation of stress-related neural interactions \citep{kim2025unlocking, li2024bigger}. The CNN encoder extracts hierarchical spatial representations from TV-PDC maps to characterize localized and distributed stress-related cerebral connectivity patterns \citep{bagherzadeh2022recognition, dosovitskiy2020image}. CNNs such as VGG16, ResNet50, DenseNet121, EfficientNet-V2, and ViT models such as LAION-CLIP-ViT-L14, OpenAI CLIP ViT-Large-Patch14 (CLIP-ViT-L/14), LAION-CLIP-ViT-H14, OpenAI CLIP ViT-Base-Patch16 (CLIP-ViT-B/16), and OpenAI CLIP ViT-Base-Patch32 (CLIP-ViT-B/32) are implemented in this research. To reduce redundancy and enhance generalizability, Principal Component Analysis (PCA) is applied to the extracted embeddings, retaining the most informative components \citep{salam2026improving}. Finally, the reduced feature representations are classified using lightweight ML methods such as Support Vector Machines (SVM), Decision Tree (DT), Gradient Boost (GB), Naïve Bayes (NB), K-Nearest Neighbour (KNN), Extreme Gradient Boost (XGB), Random Forest (RF) and Logistic Regression (LR) for mental stress. This enables hybrid modeling of dynamic EEG connectivity while maintaining interpretability and computational efficiency. All experiments were conducted using MATLAB (R2022b) and Python deep learning libraries on a workstation equipped with an Intel Core i7 processor and NVIDIA GPU.
	
	\subsection{Time-varying effective connectivity}
	
	Time-varying effective connectivity characterizes the dynamic, directed causal influences among distributed brain regions, enabling the examination of how neural information flow evolves in response to varying psychological stress levels \citep{wang2023using}. Partial Directed Coherence (PDC) is a Granger Causality-based effective connectivity measure widely used in EEG analysis to quantify directional information flow between cerebral cortices \citep{acharya2025neural}.
	PDC estimates a Multivariate Autoregressive (MVAR) model of multichannel EEG signals to quantify the magnitude and direction of frequency-specific causal interactions between electrodes, enabling principled characterization of dynamic brain connectivity patterns \citep{acharya2025neural}. Specifically, PDC enables the investigation of the underlying effective connectivity and information transfer within the brain, shedding light on the cognitive processes of neural networks. PDC can be calculated by following the mathematical expression \citep{acharya2025neural},
	
	\begin{align}
		\mathrm{PDC}_{i,j}(f) &=
		\frac{A_{i,j}(f)}{\sqrt{\sum_{m=1}^{M}\left|A_{m,j}(f)\right|^{2}}} \\
		&=\frac{A_{i,j}(f)}{\sqrt{a_j^{*}(f)\,a_j(f)}}.
	\end{align}
	Hence, the Time-Varying PDC could be calculated by employing \citep{acharya2025neural, baccala2001partial},
	\begin{align}
		\mathrm{PDC}_{i,j}(f,t)
		&= \frac{\left|A_{i,j}(f,t)\right|}
		{\sqrt{\sum_{m=1}^{M}\left|A_{m,j}(f,t)\right|^{2}}} \\
		&= \frac{\left|A_{i,j}(f,t)\right|}
		{\sqrt{a_j^{*}(f,t)\,a_j(f,t)}} .
	\end{align}
	In this formulation, $A_{i,j}$ denotes the $(i,j)^{\text{th}}$ element of the inverse transfer matrix $A(f)$, while $a_j(f)$ represents the $j^{\text{th}}$ column of $A(f)$. $\mathrm{PDC}_{i,j}$ is normalized to express the proportion of directed information outflow from channel $j \rightarrow i$ at frequency $f$ relative to the total outflow from channel $j$ to all connected channels. While source-level connectivity offers enhanced anatomical interpretability, sensor-level PDC is commonly used to capture directed information flow between EEG electrodes without the need for inverse modeling \citep{kaminski2014directed}. Moreover, source localization introduces additional assumptions and uncertainties related to head and forward modeling. Consequently, sensor-level PDC is employed to analyze dynamic interactions across EEG channels \citep{kaminski2014directed, blinowska2011review}. Under the PDC formulation, channels with consistently higher outgoing PDC values are interpreted as sources of information flow, whereas channels receiving stronger normalized influences are regarded as targets within the network.
	
	\begin{algorithm}[h]
		\caption{Time-Varying Partial Directed Coherence (PDC) Computation}
		\label{alg:tvpdc}
		\begin{algorithmic}[1]
			
			\Require EEG signal $\mathbf{X} \in \mathbb{R}^{T \times N}$, sampling rate $F_s$, model order $P$, window size $L$, overlap $O$, number of frequency bins $F$
			\Ensure Time-varying PDC tensor $\mathcal{P} \in \mathbb{R}^{N \times N \times W}$, where $W$ is the number of windows
			
			\State Segment $\mathbf{X}$ into $W$ overlapping windows of length $L$ and overlap $O$
			
			\For{$t \gets 1$ to $W$}
			\State $\mathbf{X}(t)$ $\gets$ EEG segment of shape $L \times N$
			\State Estimate MVAR coefficients $\{A_p(t)\}_{p=1}^{P}$ from $\mathbf{X}(t)$
			
			\For{$m \gets 0$ to $F-1$}
			\State $f_m \gets \dfrac{m F_s}{F}$
			\State $A(f_m,t) \gets I - \sum_{p=1}^{P} A_p(t) e^{-j2\pi f_m p / F_s}$
			
			\For{$j \gets 1$ to $N$}
			\State $D_j(f_m,t) \gets \sum_{k=1}^{N} |A_{k j}(f_m,t)|^2$
			
			\For{$i \gets 1$ to $N$}
			\If{$D_j(f_m,t) > 0$}
			\State $\mathrm{PDC}_{ij}(f_m,t) \gets 
			\dfrac{|A_{ij}(f_m,t)|^2}{D_j(f_m,t)}$
			\Else
			\State $\mathrm{PDC}_{ij}(f_m,t) \gets 0$
			\EndIf
			\EndFor
			
			\EndFor
			\EndFor
			
			\State $\mathcal{P}(:,:,t) \gets \dfrac{1}{|B|} 
			\sum_{m \in B} \mathrm{PDC}(:,:,f_m,t)$ 
			\Comment{Band-averaged over target band $B$}
			
			\EndFor
			
			\State \Return $\mathcal{P}$ \Comment{$\mathcal{P}$: column $j$ sums outflows from source $j$; $\mathrm{PDC}_{i,j}(f_m,t)$ quantifies $j \rightarrow i$}
			
		\end{algorithmic}
	\end{algorithm}
	The TV-PDC analysis was conducted using a novel connectivity framework developed in MATLAB (R2022b). The connectivity estimation consisted of the following stages, as shown in \hyperref[alg:tvpdc]{Algorithm~\ref*{alg:tvpdc}}, to extract TV-PDC features from multichannel EEG signals across all frequency bands. The initial stage involves loading the EEG recordings by specifying the number of EEG channels and key analytical parameters, which include window length, step size, and sampling frequency. In this study, all 32 EEG channels were selected from the MAT stimulus of the SAM 40 dataset \citep{ghosh2022sam}. A window length of 5 seconds and a step size of 2 seconds were employed, along with a fixed sampling frequency of 128 Hz.
	
	Following the EEG data loading and parameter definition, the EEG signals were segmented into concise temporal segments using the Short-Time Windowing (STW) technique \citep{liu2023multiwavelet, bagherzadeh2022recognition}. This segmentation enables the analysis of temporal brain dynamics by monitoring the evolution of connectivity patterns over time \citep{liu2023multiwavelet}. To capture these temporal variations, short 5-second windows were employed in this study. Consistently, prior studies have also successfully applied STW with a window length of 5 seconds \citep{acharya2025neural, bagherzadeh2022recognition}. An overlapping step size of 2 seconds was adopted to ensure smooth temporal transitions between successive segments, which is approximately half the window duration. Hence, using a 5-second window with a 2-second step size over the 25-second MAT trial yielded seven temporal segments (T1–T7) for each EEG trial.
	
	Once segmentation was performed, individual EEG segments were extracted by computing their corresponding start and end indices. The next critical step involves determining the model order, which specifies the number of past time points of each EEG channel used to predict its future values \citep{aho2014model}. Appropriate selection of the model order is vital to achieve a balance between accurately modeling underlying neural dynamics while avoiding overfitting or underfitting \citep{privalsky2020multivariate}. A model order 5 was selected based on optimal model order estimation using multiple statistical criteria, including Akaike’s Final Prediction Error (FPE), Hannan–Quinn criterion (HQ), Akaike’s Information Criterion (AIC), and Schwarz–Bayes Criterion (SBC). These criteria were computed using the ARfit toolbox in conjunction with the Source Information Flow Toolbox (SIFT) extension within EEGLAB \citep{aho2014model, cui2008bsmart, mullen2010source}. The mathematical formulations of AIC, SBC, FPE, and HQ are detailed in \citep{mullen2010source}.
	
	After identifying the optimal model order, the next step was to fit an MVAR model to the segmented EEG data. This step models directional causal interactions among EEG channels by estimating their time-varying relationships. In this study, the Yule–Walker method was used to compute the MVAR coefficients \citep{khan2023novel}. Model fitting refers to estimating parameters that characterize how past channel activities influence the current EEG signals. The stability of the estimated MVAR models was verified before connectivity estimation to ensure a valid connectivity analysis. Given the sampling frequency, the MVAR-based transfer function was evaluated in the frequency domain, and the resulting frequency indices were used to compute EEG-specific PDC estimates for the canonical EEG bands listed in \autoref{tab:eeg_spectrum}. Frequency-domain responses were then computed for each pair of EEG channels within these bands to quantify directed interactions across EEG frequency bands. This computation quantifies the directional influence of one channel on another at specific frequencies, which is a core component of the TV-PDC analysis.
	
	\subsection{Deep embeddings extraction}
	
	The TV-PDC maps represent frequency-specific information flow matrices that encode pairwise causal interactions among EEG channels. The temporal sequence of these matrices captures the dynamic patterns of large-scale effective brain networks associated with mental stress processing across successive time windows. Connectivity matrices encode spatial interaction patterns between EEG channels and can therefore be represented as structured images, enabling convolutional and transformer architectures to learn relational representations of directed brain networks \citep{wang2023using, bagherzadeh2022recognition}. CNNs are particularly effective for TV-PDC maps due to their capacity to explore local spatial correlations, connectivity patterns, and multiscale structural regularities within directed matrices through hierarchical convolutional receptive fields \citep{bagherzadeh2022recognition}. These properties enable CNNs to capture regionally localized patterns of directed information flow and spatially clustered causal interactions among EEG channels \citep{bagherzadeh2022recognition}. In parallel, ViT models are employed to model global dependencies and long-range inter-channel relationships across the connectivity maps using self-attention mechanisms \citep{dosovitskiy2020image}. This global modeling capability is critical for characterizing the distributed, network-wide connectivity patterns that may not be fully captured by locality-biased convolutional operations \citep{dosovitskiy2020image}. Consequently, the complementary integration of CNNs and ViTs facilitates simultaneous learning of localized connectivity patterns and global brain network dependencies, which are inherent in TV-PDC representations.
	
	Among the CNNs explored in this research, VGG16 is a deep CNN composed of sequential stacks of small 3×3 convolutional filters that enable hierarchical feature extraction with uniform architectural depth \citep{khan2025transfer}. It is employed in this study to provide a stable and transparent architecture for extracting spatial connectivity patterns from TV-PDC maps, while facilitating comparative analysis with more advanced architectures. Similarly, ResNet50 is a deep residual CNN that incorporates identity-based skip connections to enable effective training of substantially deeper models \citep{khan2025transfer}. Its inclusion is motivated by its ability to extract high-level features from complex directed connectivity matrices, making it an appropriate choice for sophisticated connectivity patterns in TV-PDC maps. Consequently, EfficientNet-V2 is a compound scaled convolutional architecture that jointly optimizes network depth, width, and input resolution for improved parameter efficiency and performance. It is adopted to enable robust multiscale feature extraction from high-dimensional TV-PDC maps while maintaining computational efficiency and reduced model complexity. Among the ViT models, LAION-CLIP-ViT-L14 is a large-scale ViT model, pretrained using contrastive image or text objectives with a patch size of 14×14 \citep{li2024bigger}. It is employed to extract high-capacity global embeddings that capture long-range inter-channel dependencies and distributed connectivity patterns across TV-PDC maps. Likewise, CLIP-ViT-L/14 is another ViT model, pretrained on large-scale multi-modal data, producing semantically rich and transferable representations \citep{abbasi2024language}. This model is used for its strong global self-attention capability, which enables effective modeling of network-wide connectivity organization beyond localized spatial patterns. Furthermore, LAION-CLIP-ViT-H14 is a high-capacity ViT model with increased depth and attention heads, designed to capture complex relational structures \citep{tran2026brittleness}. This model is included to extract higher-order global dependencies and directed information-flow patterns from TV-PDC connectivity maps. Subsequently, CLIP-ViT-B/16 is applied, which is a compact ViT architecture with moderate patch granularity \citep{kim2025unlocking}. This model offers a balance between local detail and global context \citep{kim2025unlocking}. It is employed to capture fine-grained spatial variations in information flow patterns while retaining global self-attention for holistic network modeling. Lastly, CLIP-ViT-B/32 is a lightweight ViT model applied in this research, which includes patch tokenization of size 32×32 and reduced computational complexity \citep{abbasi2024language}. This model is utilized to emphasize global connectivity and electrode interactions within TV-PDC maps, complementing a finer-resolution ViT variant.
	
	Following deep feature extraction, PCA was applied to the extracted deep feature vectors, retaining principal components that preserved 95\% of the cumulative explained variance. PCA serves to compact the feature representations derived from CNNs and ViTs, while mitigating redundancy across correlated features \citep{salam2026improving}. PCA also improves the stability of subsequent ML classifiers, particularly in high feature-to-sample ratio settings commonly used in EEG connectivity analysis \citep{salam2026improving}. To strictly avoid data leakage, PCA is applied within the training folds only, where the projection matrix is learned exclusively from the training data during cross-validation. The learned PCA transformation is then applied to the corresponding test data, ensuring that no information from unseen samples influences the feature space construction. This fold-wise PCA strategy preserves the integrity of the evaluation protocol while enabling efficient dimensionality reduction of TV-PDC-based deep embeddings without compromising statistical validity \citep{salam2026improving}.
	
	\subsection{Machine learning classifiers}
	
	In this experiment, eight ML classifiers such as SVM \citep{attallah2020effective}, DT \citep{rahman2022detection}, GB \citep{rahman2022detection}, NB \citep{katmah2021review}, KNN \citep{shashidhar2023eeg}, RF \citep{nasteski2017overview}, XGB \citep{rahman2022detection} and LR \citep{katmah2021review} were employed for mental stress classification. These classifiers were selected to leverage their diverse learning mechanisms, robustness in operating high-dimensional feature spaces, and their suitability for limited EEG data.
	
	SVM and DT have been previously used in an EEG-based study for decoding motor imagery utilizing PDC connectivity features \citep{rahman2022detection}. In this work, following the deep feature extraction from TV-PDC maps using CNN and ViT models and subsequent dimensionality reduction via PCA, SVM is used to learn discriminative decision boundaries within the resulting latent feature space, while DT provides an interpretable hierarchical structure for identifying stress-related patterns embedded in the learned connectivity representations \citep{attallah2020effective, rahman2022detection}.
	
	Instance-based and probabilistic classifiers, such as KNN and NB, were incorporated to assess local neighborhood relationships and probabilistic class separability within the learned TV-PDC feature representations \citep{shashidhar2023eeg, katmah2021review}. KNN captures similarity-driven patterns among the latent representations without imposing parametric assumptions. In contrast, NB provides a computationally efficient baseline by modeling class-conditional probability distributions under a conditional independence assumption \citep{shashidhar2023eeg, katmah2021review}. 
	
	Ensemble classifiers, such as RF, GB, and XGB, were employed to model non-linear and higher-order relationships within the PCA-reduced latent representations derived from the deep features of TV-PDC maps \citep{rahman2022detection}. Through aggregating multiple weak learners across bagging and boosting strategies, these ensemble methods can improve robustness and reduce overfitting when operating on deep connectivity embeddings \citep{nasteski2017overview, rahman2022detection}. LR was additionally included as a lightweight classifier to provide probabilistic interpretability of stress-related decision boundaries within the learned TV-PDC-derived latent feature space \citep{katmah2021review}.
	
	To evaluate the classification performance, standard metrics such as accuracy, precision, recall, and F1-score were employed. Precision quantifies the ratio of correctly identified positive instances among all predicted positives, reflecting the reliability of stress detection \citep{nasteski2017overview}. Recall represents the ratio of true positive instances correctly identified among all actual positives, indicating the sensitivity of the classifier \citep{rahman2022detection}. The F1-score is defined as the harmonic mean of precision and recall, which provides a balanced performance measure \citep{nasteski2017overview}. Accuracy denotes the ratio of correctly classified instances, including both true positives and true negatives, to the total number of predictions \citep{rahman2022detection}. The formulations of performance metrics in Eqs.~(\ref{eq:precision})--(\ref{eq:accuracy}) are provided below \citep{nasteski2017overview, rahman2022detection}.
	\begin{gather}
		\mathrm{Precision} = \frac{T_p}{T_p + F_p} \label{eq:precision} \\
		\mathrm{Recall} = \frac{T_p}{T_p + F_n} \label{eq:recall} \\
		\mathrm{F1\ score} = 2 \times 
		\frac{\mathrm{Precision} \times \mathrm{Recall}}
		{\mathrm{Precision} + \mathrm{Recall}} \label{eq:f1} \\
		\mathrm{Accuracy} = 
		\frac{T_p + T_n}{T_p + T_n + F_p + F_n} \label{eq:accuracy}
	\end{gather}
	Where $T_p$ represents the true positives, $T_n$ represents the true negatives, $F_p$ represents the false positives, and $F_n$ represents the false negatives.
	
	\section{Classification results}
	
	This section presents a comprehensive evaluation of the stress-classification results obtained using the NeuroStrata framework for EEG frequency-specific TV-PDC features. Classification performance is analyzed across multiple deep feature extractors and lightweight ML classifiers using accuracy, precision, recall, and F1-score to assess the discriminative capability of directed EEG connectivity patterns under varying stress levels. Furthermore, a subject-wise 10-fold cross-validation strategy was employed, in which all temporal segments from a given participant were assigned to the same fold. This ensured that EEG recordings from the same participant were not simultaneously present in both the training and testing sets, thereby preventing subject-wise data leakage and enabling robust evaluation of model generalization performance.
	
	\begin{figure*}[t]
		\centering
		
		\begin{subfigure}[h]{0.48\textwidth}
			\centering
			\includegraphics[width=\linewidth]{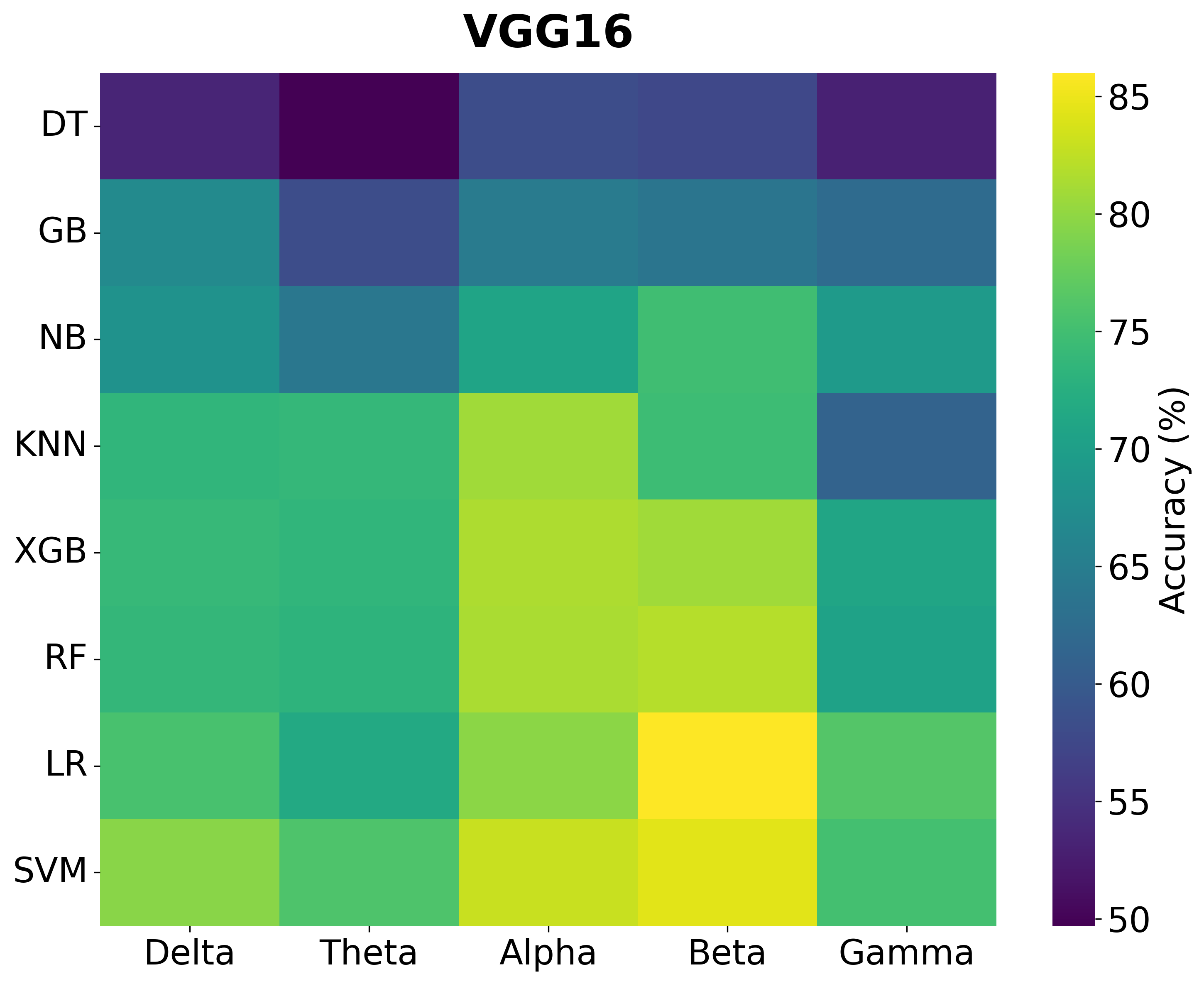}
		\end{subfigure}
		\hfill
		\begin{subfigure}[h]{0.48\textwidth}
			\centering
			\includegraphics[width=\linewidth]{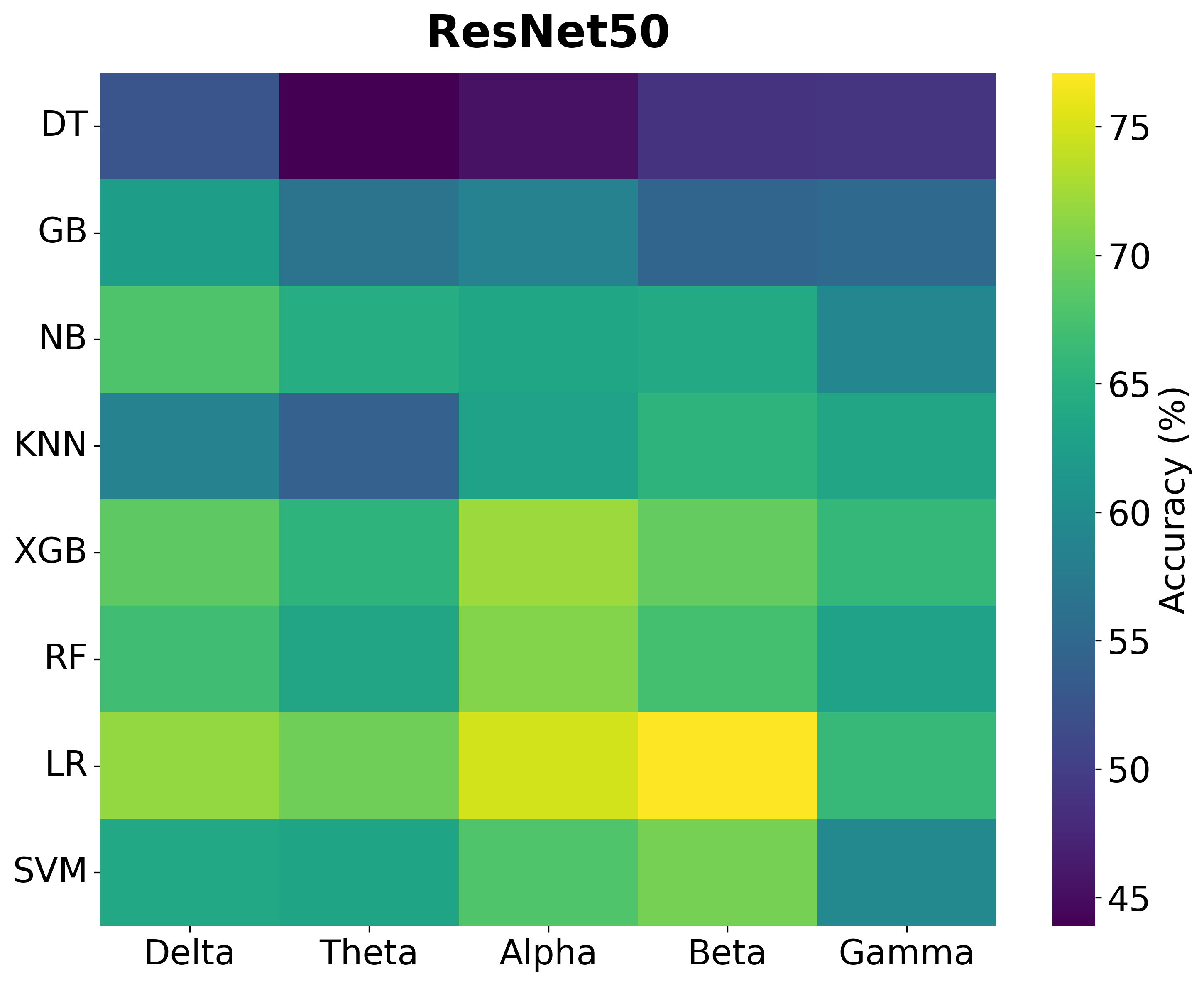}
		\end{subfigure}
		
		\vspace{2mm}
		
		\begin{subfigure}[h]{0.48\textwidth}
			\centering
			\includegraphics[width=\linewidth]{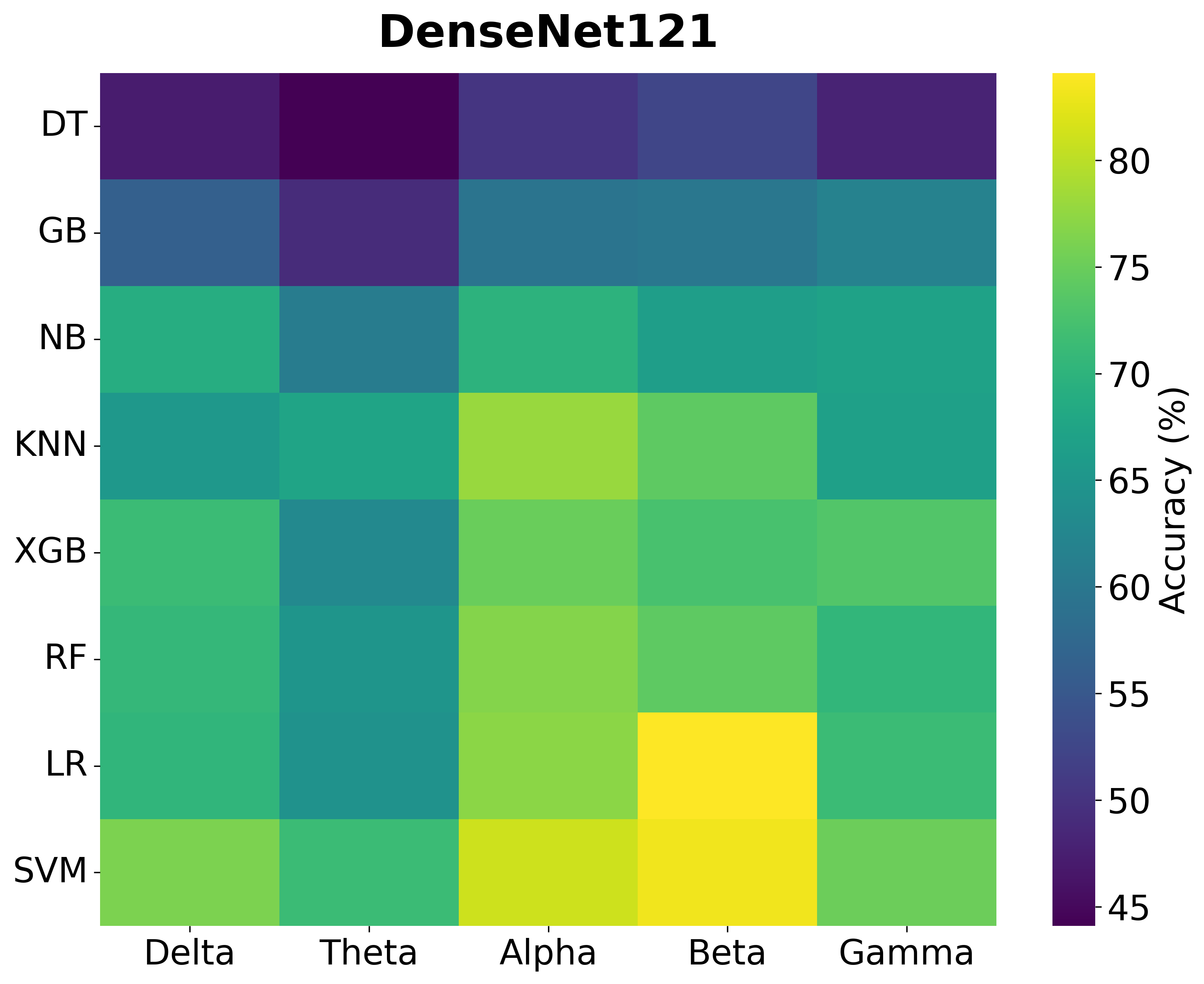}
		\end{subfigure}
		\hfill
		\begin{subfigure}[h]{0.48\textwidth}
			\centering
			\includegraphics[width=\linewidth]{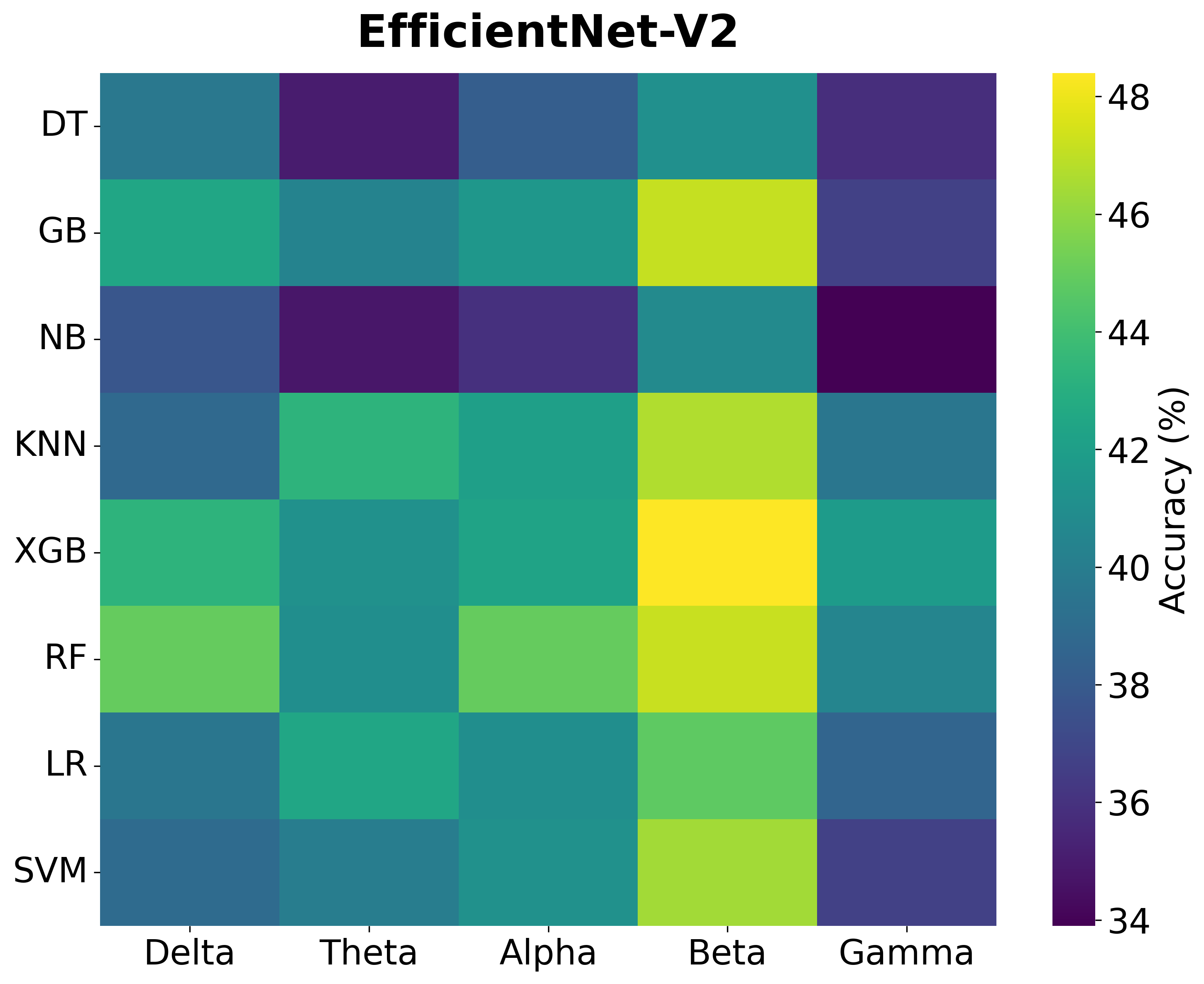}
		\end{subfigure}
		
		\caption{Classification results obtained using four CNN backbones (VGG16, ResNet50, DenseNet121, and EfficientNet-V2).}
		\label{fig:cnn_backbones}
		
	\end{figure*}
	
	\begin{figure*}[!t]
		\centering
		
		\newcommand{\vitpanelheight}{0.27\textheight}
		
		\includegraphics[height=\vitpanelheight]{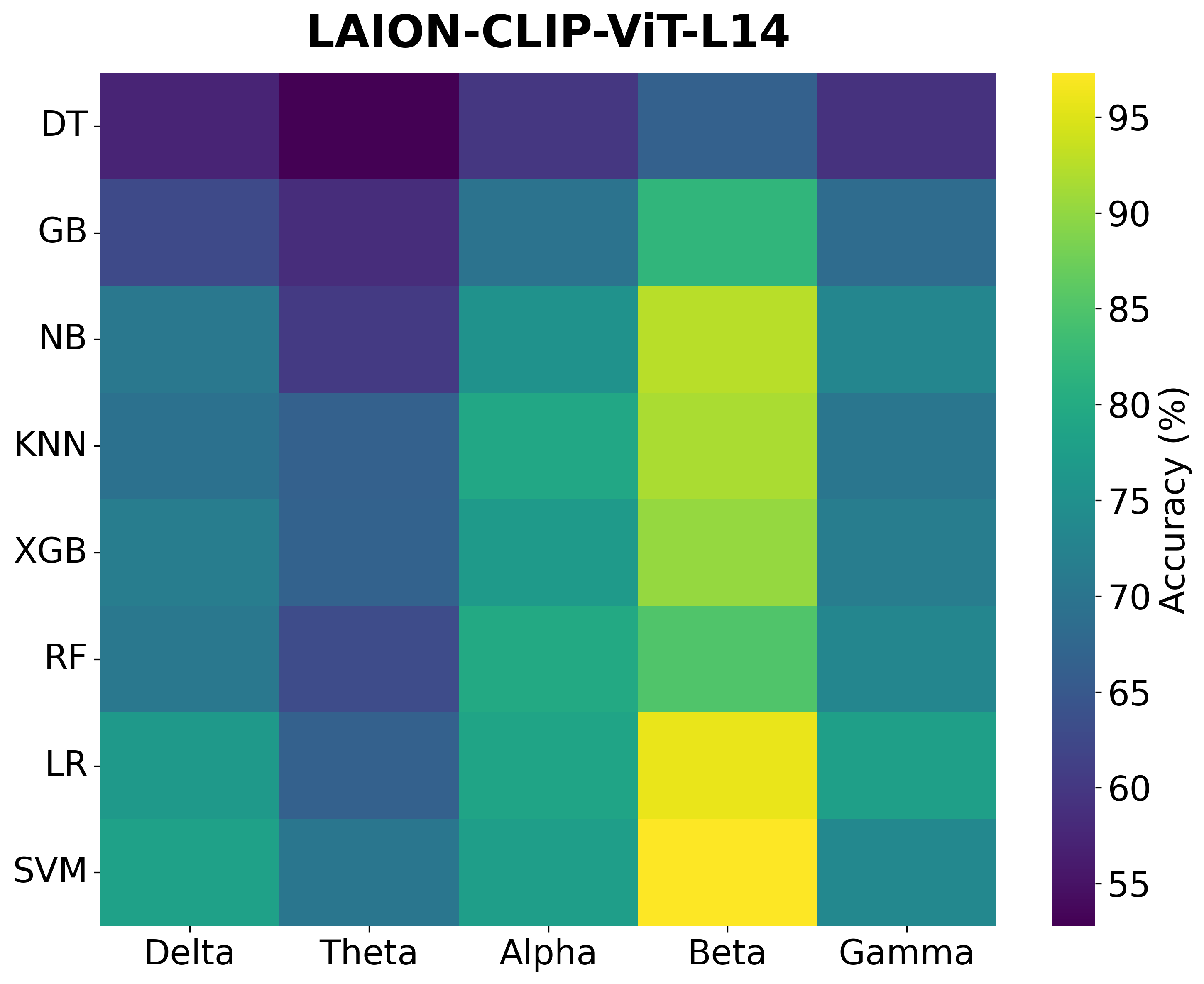}
		\hfill
		\includegraphics[height=\vitpanelheight]{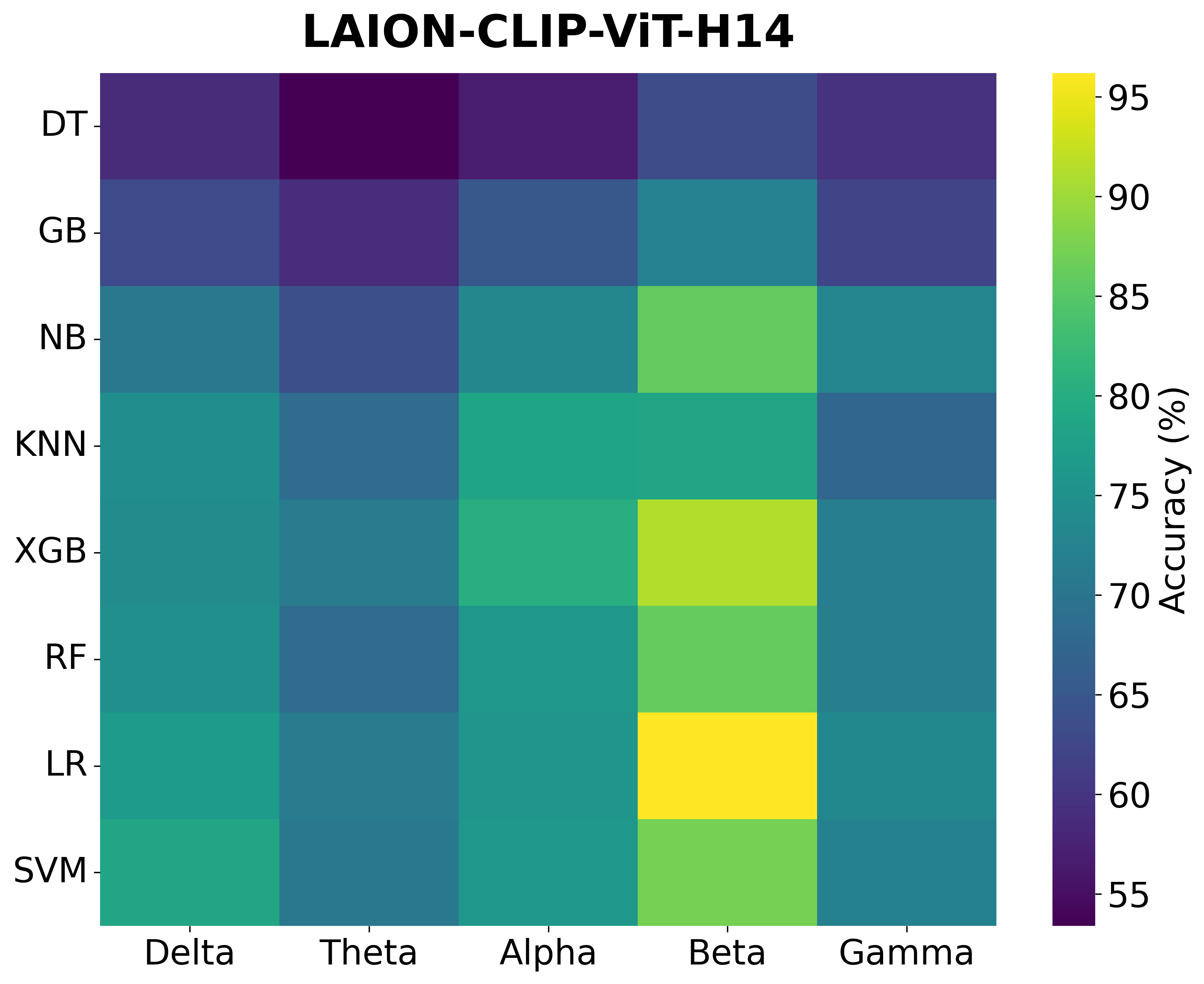}
		
		\vspace{1.5mm}
		
		\includegraphics[height=\vitpanelheight]{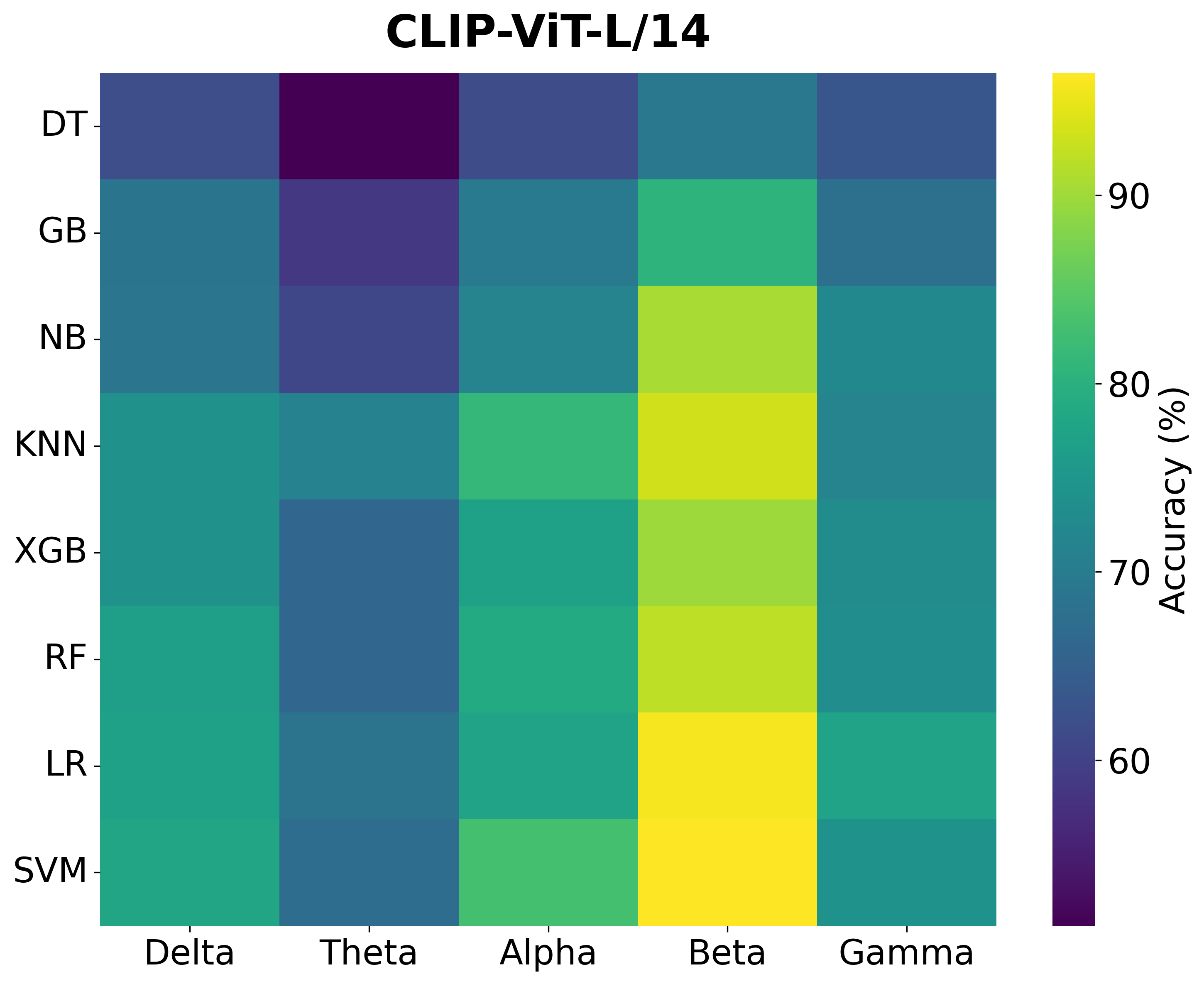}
		\hfill
		\includegraphics[height=\vitpanelheight]{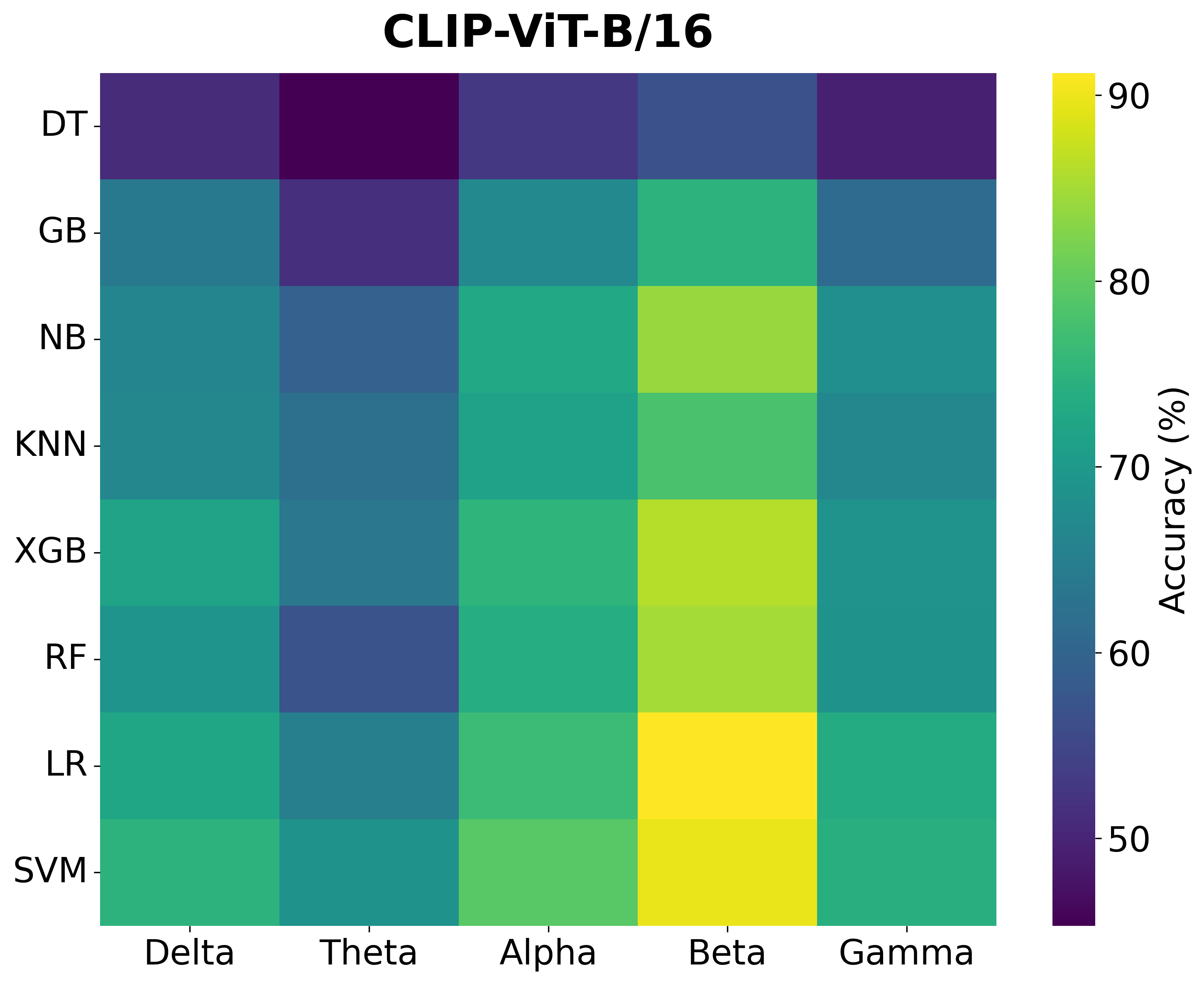}
		
		\vspace{1.5mm}
		
		\makebox[\textwidth][c]{%
			\includegraphics[height=\vitpanelheight]{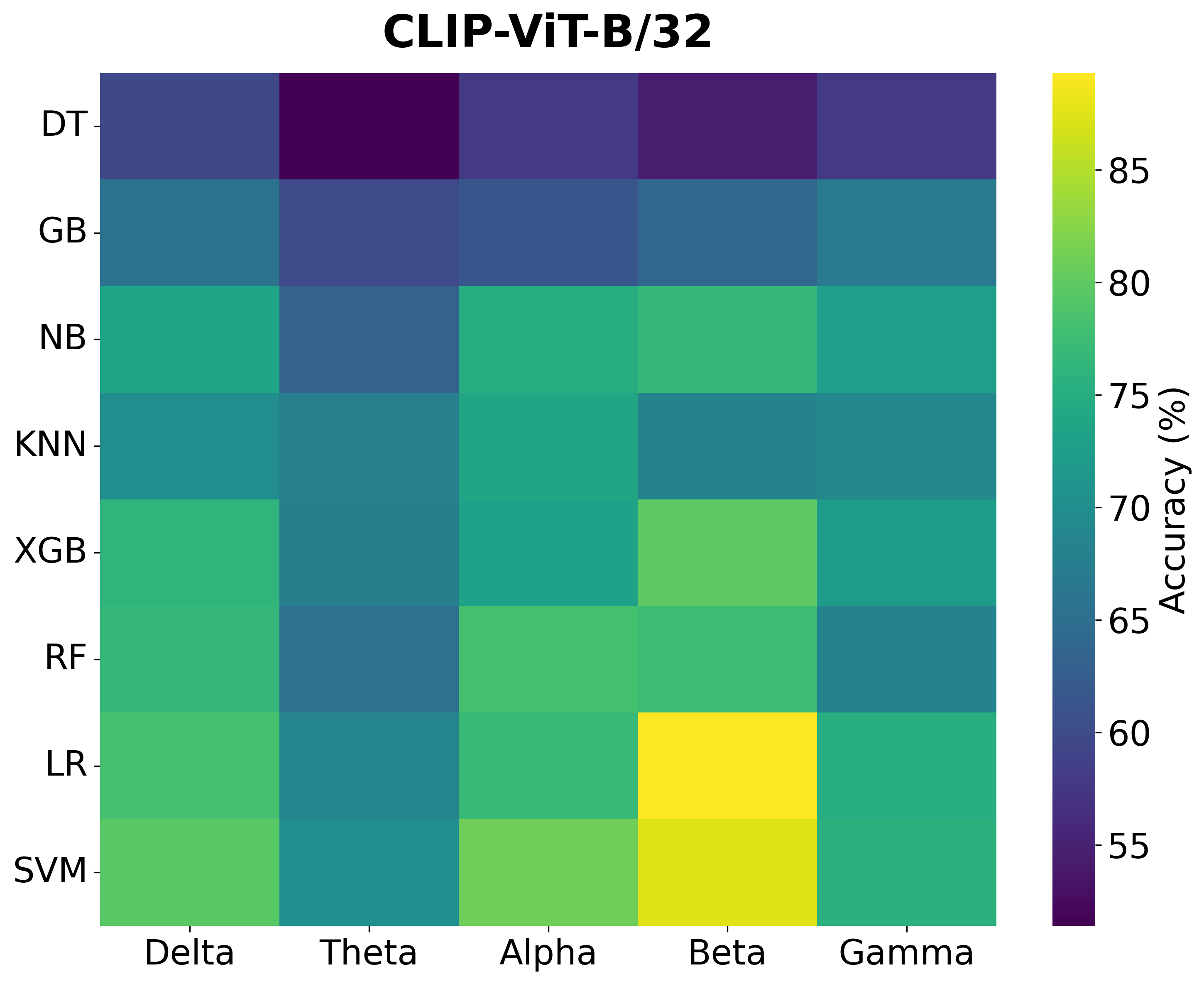}
		}
		
		\caption{Classification results obtained using five ViT-based backbones (LAION-CLIP-ViT-L14, LAION-CLIP-ViT-H14, CLIP-ViT-L/14, CLIP-ViT-B/16, and CLIP-ViT-B/32).}
		\label{fig:vit_backbones}
	\end{figure*}
	
	From \autoref{fig:cnn_backbones} and \autoref{fig:vit_backbones}, delta-PDC demonstrates comparatively lower yet stable classification performance across all DL-ML combinations. For CLIP-ViT-B/16, SVM achieved an accuracy of 74.8\%, precision 75.7\%, recall 74.8\%, and F1 score 74.8\% while outperforming RF, which produced 69.1\% accuracy and XGB with 72.0\% accuracy. A similar trend is observed with CLIP-ViT-B/32, where SVM achieved 78.1\% accuracy and 78.1\% F1, exceeding RF (76.7\%) and XGB (76.1\%). Alongside stronger embeddings such as CLIP-ViT-L/14, delta-PDC performance improved further, with SVM attaining 77.8\% accuracy and 77.8\% F1, while RF and XGB achieved 76.6\% and 73.9\% accuracy, respectively. Among CNN backbones, EfficientNet-V2 achieved an SVM accuracy of 78.5\%, whereas ResNet-50 and VGG-16 reported lower accuracies of 78.1\% and 63.8\%, respectively. From the results, delta-PDC provides consistent but limited discriminative power, specifically when compared with higher EEG bands.
	
	On the other hand, theta-PDC outperforms delta-PDC across all DL backbones. In \autoref{fig:vit_backbones} for CLIP-ViT-B/16, SVM achieved 68.7\% accuracy, 66.9\% precision, 67.8\% recall, and 67.8\% F1, while RF and XGB reached 57.0\% and 63.5\% accuracy, respectively. Along with CLIP-ViT-B/32, SVM accuracy increased to 70.2\% and F1 score to 69.8\%. Notably, CLIP-ViT-L/14 produced further gains, where SVM achieved 67.3\% accuracy and 67.3\% F1, while RF obtained 66.1\% accuracy. Among the CNN models in \autoref{fig:cnn_backbones}, DenseNet121 achieved the highest Theta-PDC performance, with an SVM accuracy of 71.4\%, followed by EfficientNet-V2 at 70.5\% and ResNet50 at 70.3\%. These results indicate that theta-PDC captures stress-related directional interactions more effectively than delta-PDC, though it remains inferior to alpha- PDC and beta-PDC features.
	
	Alpha-PDC consistently delivers strong and reliable classification performance across all DL-ML configurations. For CLIP-ViT-B/16 in \autoref{fig:vit_backbones}, SVM achieved 79.2\% accuracy, 80.6\% precision, 79.2\% recall, and 79.2\% F1 score, clearly surpassing RF, which yielded 73.7\% accuracy, and XGB, with 75.2\% accuracy. Similar improvements have been observed with CLIP-ViT-B/32, where SVM achieved 81.0\% accuracy and 81.0\% F1 score. The best alpha-PDC performance is observed with CLIP-ViTL/14, where SVM attained 83.0\% accuracy, 83.8\% precision, 83.0\% recall, and 83.0\% F1 score, demonstrating effective window-level stress discrimination using pre-trained deep representations. RF achieved 78.8\% accuracy, and XGB 77.1\%. Among the CNN-based models in \autoref{fig:cnn_backbones}, enseNet121 and EfficientNet-V2 achieved comparable peak alpha-PDC performance, each reaching 81.0\% accuracy with SVM, while ResNet50 and VGG16 reported 77.6\% and 68.0\% accuracy, respectively. These findings confirm that alpha-based directional connectivity provides a robust representation of stress-related neural dynamics.
	
	\begin{figure*}[t]
		\centering
		
		\begin{subfigure}[t]{0.49\textwidth}
			\centering
			\includegraphics[width=\linewidth]{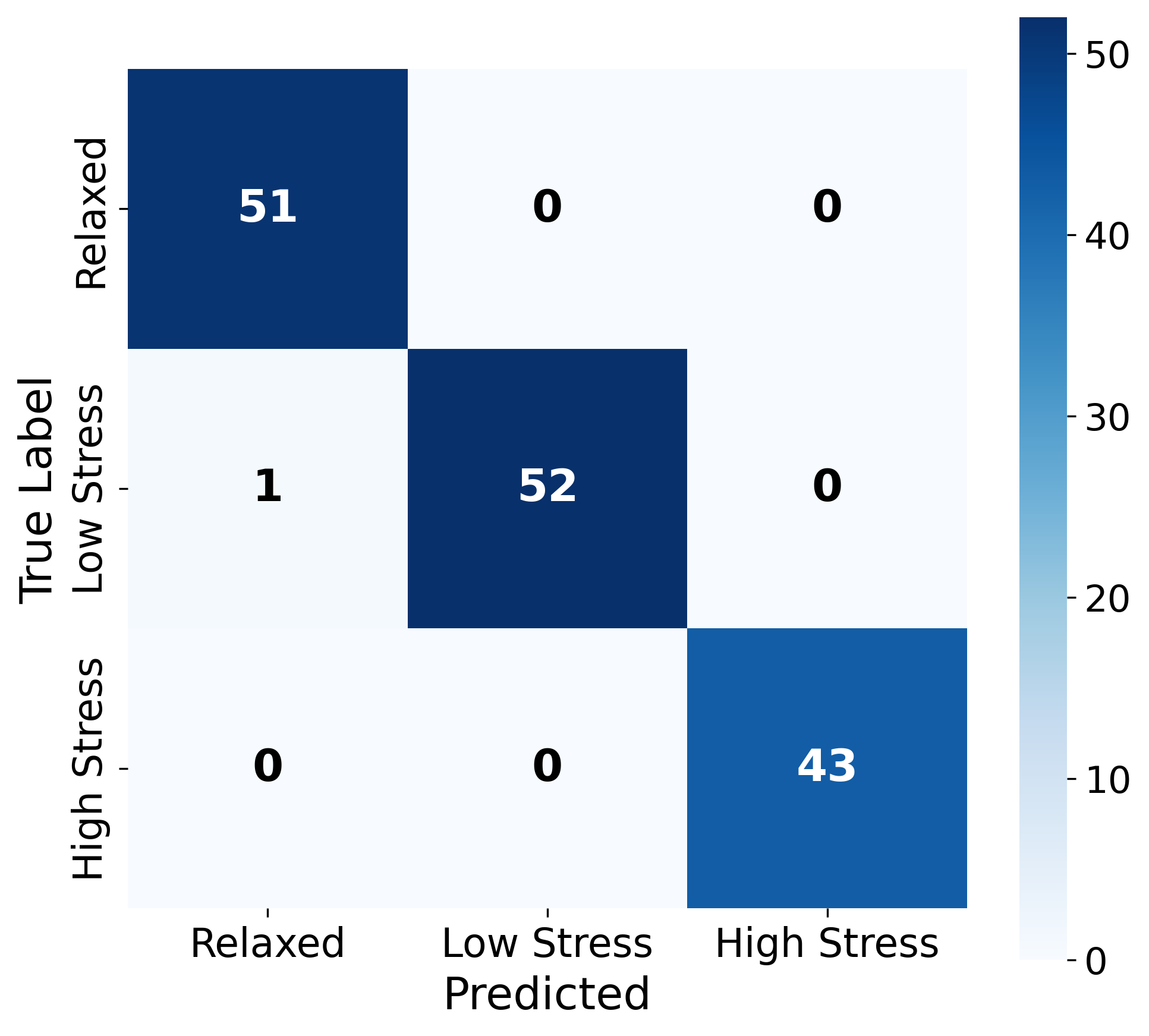}
			\caption{Confusion matrix of raw counts using beta-PDC with LAION-CLIP-ViT-L14 and SVM.}
			\label{fig:cm_beta_laionl14_svm_counts}
		\end{subfigure}
		\hfill
		\begin{subfigure}[t]{0.49\textwidth}
			\centering
			\includegraphics[width=\linewidth]{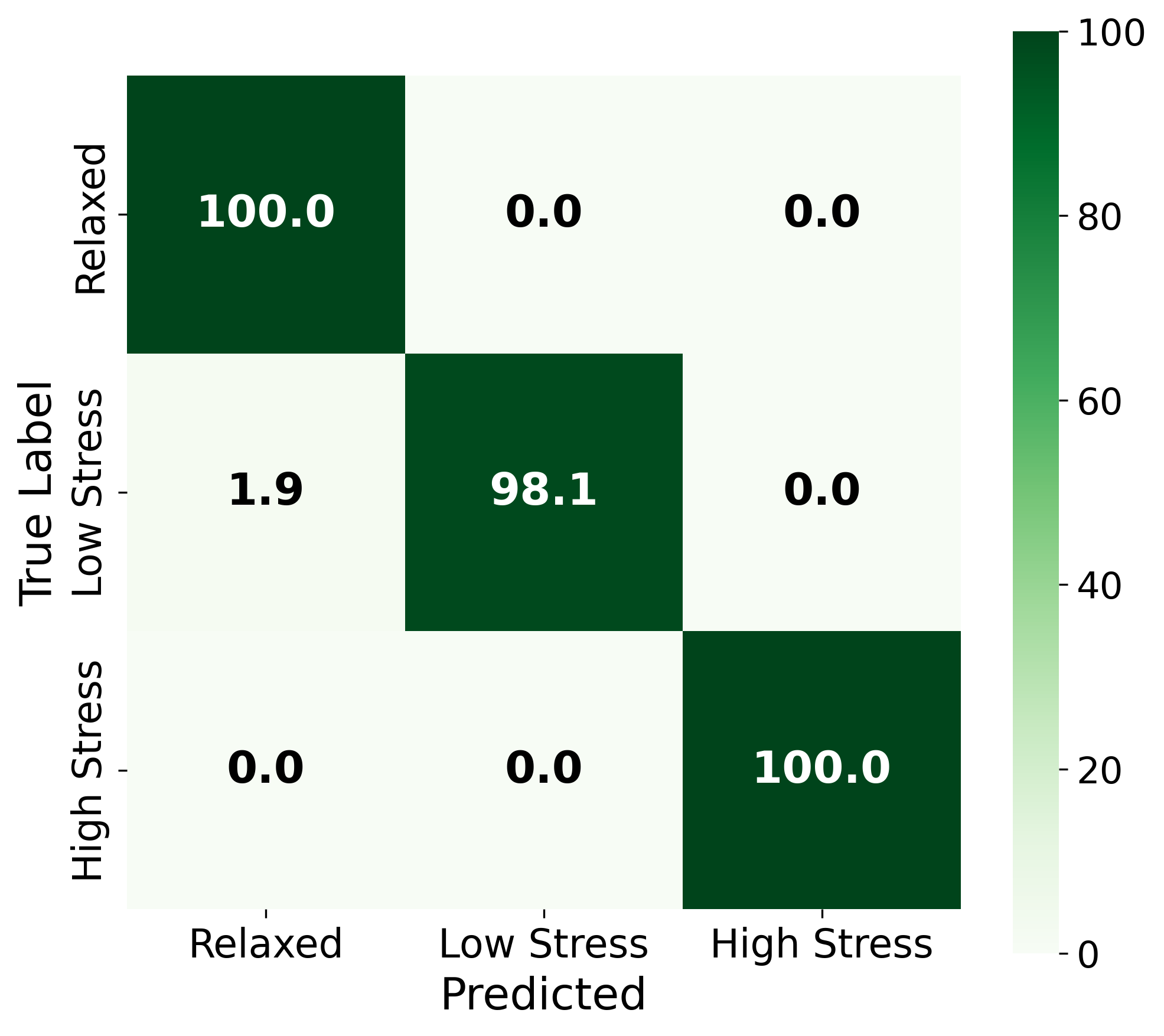}
			\caption{Confusion matrix of percentages using beta-PDC with LAION-CLIP-ViT-L14 and SVM.}
			\label{fig:cm_beta_laionl14_svm_percent}
		\end{subfigure}
		
		\caption{Confusion matrices of raw counts and percentages for beta-PDC using LAION-CLIP-ViT-L14 with SVM.}
		\label{fig:cm_beta_2models}
	\end{figure*}
	
	Beta-PDC exhibits the most discriminative behavior across the evaluated bands for all DL feature backbones. For CLIP-ViT-B/16 in \autoref{fig:vit_backbones}, LR achieved strong performance, with 91.2\% accuracy, 91.3\% precision, 91.2\% recall, 91.2\% F1, exceeding SVM, which yielded 89.7\% accuracy, 89.9\% precision, 89.7\% recall, 89.7\% F1 score. A similar pattern is observed for CLIP-ViT-B/32, where LR again led with 89.3\% accuracy, followed by SVM at 87.5\% accuracy, while XGB and RF reached 79.9\% and 77.4\% accuracy, respectively. CLIP-ViT-L/14, while combined with SVM, attained 96.5\% accuracy, 96.6\% precision, 96.5\% recall, 96.5\% F1, followed by LR at 95.9\% accuracy and RF at 92.0\% accuracy. For DenseNet121  in \autoref{fig:cnn_backbones}, LR yielded 84.1\% accuracy, 83.9\% precision, 83.6\% recall, and 83.8\% F1, marginally higher than SVM (83.2\% accuracy) and substantially higher than RF (74.2\% accuracy). In contrast, EfficientNet-V2 exhibited notably lower beta-PDC scores. EfficientNet-V2 with XGB achieved 48.4\% accuracy, 49.3\% precision, 48.4\% recall, 48.4\% F1 score. The LAION-CLIPViT- H14 backbone showed strong beta-PDC separability under LR, achieving 96.5\% accuracy, 96.2\% precision, 96.2\% recall, and 96.2\% F1 score, whereas LAION-CLIP-ViT-L14 achieved its best performance under SVM, reaching 97.3\% accuracy, 98.3\% precision, 97.0\% recall, and 97.3\% F1 score. For CNN feature extractors, VGG16 achieved peak accuracies of 86.0\%, precision of 86.1\%, and recall and F1 scores of 86.0\%, particularly when combined with LR. It also achieved 84.3\% with SVM. Furthermore, ResNet50 performs well with an accuracy of 77.1\%, precision of 77.4\%, recall of 77.1\%, and F1 score of 77.1\% under LR. 
	
	Confusion matrices for beta-PDC combined with LAION-CLIP-ViT-L14 and SVM are reported in \autoref{fig:cm_beta_laionl14_svm_counts} and \autoref{fig:cm_beta_laionl14_svm_percent}, showing raw counts and percentages. The raw count matrix indicates highly accurate classification: 51 Relaxed trials were correctly identified out of 51, 51 Low Stress trials were correctly classified out of 52, and 43 High Stress trials were correctly classified out of 43, corresponding to only a single misclassification between Relaxed and Low Stress categories. This matrix reflects the performance from one representative split, whereas the reported 97.3\% accuracy corresponds to the mean across the subject-wise 10-fold cross-validation folds, and therefore, slight variation from the averaged performance is expected. The percentage confusion matrix further highlights class-specific behavior, showing 100\% accuracy for Relaxed, 98.1\% for Low Stress, and 100\% for High Stress, confirming near-perfect separability of stress states using Beta-band connectivity embeddings. The minimal confusion observed primarily between Relaxed and Low Stress suggests that beta-PDC features extracted via LAION-CLIP- ViT-L14 provide highly stable and discriminative representations, with negligible fold-level variability and strong robustness across train–test splits for 3-class stress classification.
	
	Gamma-PDC results, as shown in \autoref{fig:cnn_backbones}  and \autoref{fig:vit_backbones}, demonstrate moderate classification performance across DL-ML combinations.  For CLIP-ViT-B/16, SVM achieved 74.2\% accuracy, 75.2\% precision, 74.2\% recall, and an F1 Score of 74.2\%, outperforming RF (68.6\%) and XGB (68.8\%). A similar trend is observed for CLIP-ViT-B/32: SVM achieved 75.6\% accuracy, while RF reached 68.3\%. For CLIP-ViTL/14, SVM achieved 74.3\% accuracy, followed closely by RF at 73.2\% and XGB at 73.1\%. Among the CNN models, EfficientNet-V2 achieved 75.1\% accuracy, compared to 73.6\% for ResNet50 and 59.5\% for VGG16. Although gamma-PDC provides useful complementary information, its discriminative capability remains lower than that of alpha-PDC and beta-PDC. Based on the classification results, alpha-PDC and beta-PDC were selected for further investigation because they consistently yielded the highest classification accuracies across both CNN and ViT backbones, with beta-PDC achieving peak performance and alpha-PDC showing stable, second-best discriminative capability across multiple DL-ML pipelines.
	
	\section{Analysis of important TV-PDC features}
	
	Given XGB’s predictive capability and effectiveness in modeling complex feature interactions, it is also employed in this study to assess the relative importance of TV-PDC features. XGB provides a powerful tool for identifying the most influential temporal connectivity patterns, thereby offering valuable insights into the directed interactions that contribute most significantly to stress classification \citep{hsieh2019feature}. During model training, XGB sequentially builds an ensemble of decision trees, where each subsequent tree is designed to correct the errors of its predecessors \citep{rahman2022detection, hsieh2019feature}. Feature selection in this process is guided by a split evaluation criterion known as gain, with features that produce higher gain values being preferentially selected \citep{hsieh2019feature}. By aggregating the gain values for each feature across all trees, XGB computes a normalized importance score ranging from 0 to 1 for each feature connection \citep{hsieh2019feature}. This process is facilitated by XGB’s built-in feature importance mechanism, which ranks features by their contribution to the model’s predictive performance \citep{hsieh2019feature}. Beyond identifying critical stress-related connectivity features, this importance-scoring approach also enhances the model's interpretability by clarifying how individual temporal connections influence classification outcomes \citep{hsieh2019feature}. In this work, an importance score threshold of 0.9 was applied to emphasize the most relevant alpha- PDC and beta-PDC features. Since the alpha and beta bands consistently exhibited the highest discriminative power for stress classification, alpha-PDC and beta-PDC features were selected for importance analysis to investigate the dynamic evolution of directed neural influences within these EEG bands. These importance scores quantify the discriminative relevance of connectivity features for classification rather than their physiological strength. Accordingly, the important connections represent stress-sensitive pathways rather than the prominent information flow routes.
	
	\begin{figure*}[t]
		\centering
		
		\begin{subfigure}[t]{0.49\textwidth}
			\centering
			\includegraphics[width=\linewidth]{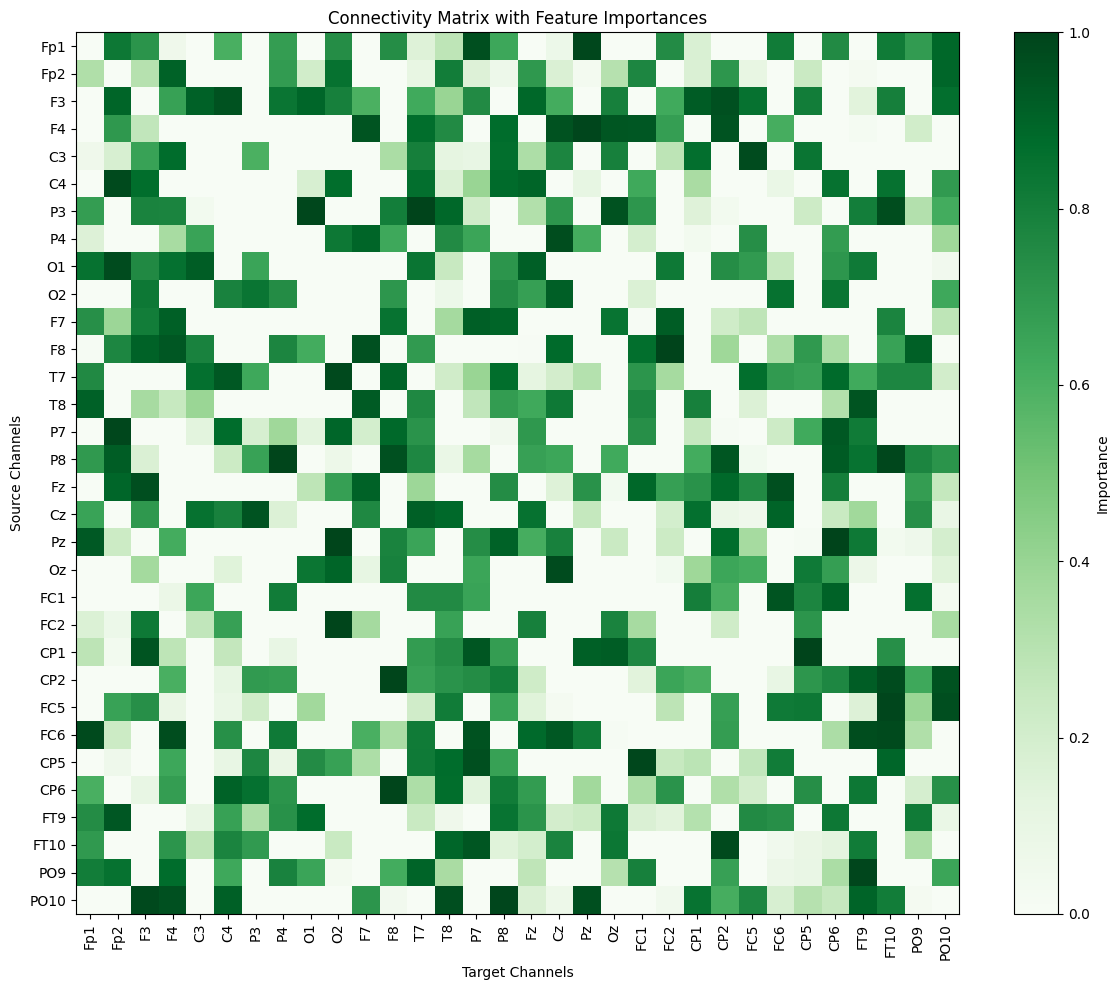}
			\caption{Feature matrix of alpha-PDC with importance scores.}
			\label{fig:alpha_feature_matrix}
		\end{subfigure}
		\hfill
		\begin{subfigure}[t]{0.49\textwidth}
			\centering
			\includegraphics[width=\linewidth]{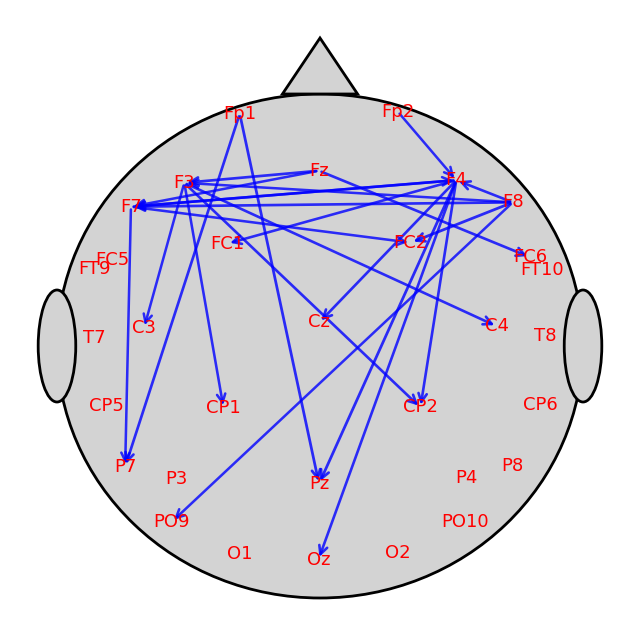}
			\caption{Directed influences of alpha-PDC above 0.9 importance score.}
			\label{fig:alpha_directed}
		\end{subfigure}
		
		\vspace{2mm}
		
		\begin{subfigure}[t]{0.49\textwidth}
			\centering
			\includegraphics[width=\linewidth]{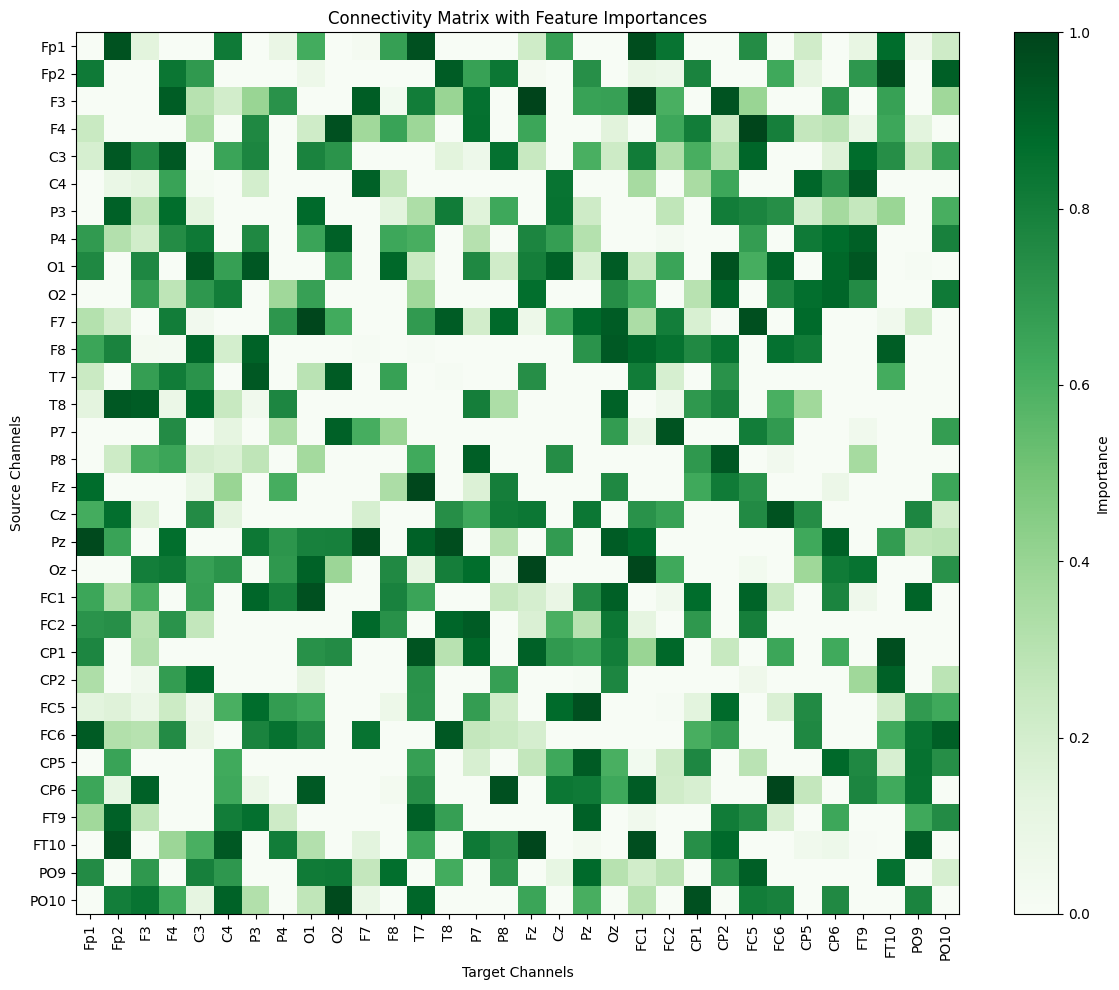}
			\caption{Feature matrix of beta-PDC with importance scores.}
			\label{fig:beta_feature_matrix}
		\end{subfigure}
		\hfill
		\begin{subfigure}[t]{0.49\textwidth}
			\centering
			\includegraphics[width=\linewidth]{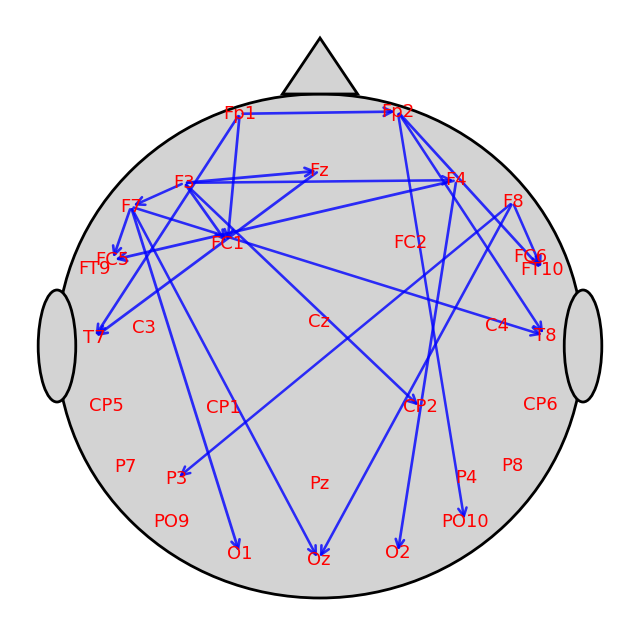}
			\caption{Directed influences of beta-PDC above 0.9 importance score.}
			\label{fig:beta_directed}
		\end{subfigure}
		
		\caption{Feature matrices with importance scores of alpha-PDC and beta-PDC connections, and EEG directed influences of alpha-PDC and beta-PDC with importance scores greater than 0.9.}
		\label{fig:pdc_feature_importance}
		
	\end{figure*}
	
	All alpha-PDC feature connections with their corresponding importance scores are illustrated in \autoref{fig:alpha_feature_matrix}, while the resulting EEG topological representation of the most influential connections is shown in \autoref{fig:alpha_directed}. The alpha-PDC connectivity pattern indicates prominent involvement of frontal and fronto-central regions, which act as major sources of directed information flow. In particular, frontal electrodes such as Fp1, Fp2, F3, F4, F7, F8, Fz, FC1, and FC2 emerge as strong drivers, exerting directed influences toward posterior regions, including parietal and occipital sites such as Pz, P7, P8, O1, O2, PO9, and PO10, while central regions primarily act as intermediate targets. This directed topology indicates an anterior-posterior information flow, reflecting top-down modulation from frontal executive regions toward posterior sensory and integrative areas during stress processing. Several long-range fronto-parietal and fronto-occipital connections are particularly pronounced, including pathways originating from F3, F4, F7, F8, and Fz toward Pz, P7, P8, O1, and O2, highlighting the role of frontal alpha activity in regulating posterior cortical dynamics. A hemispheric propensity is also evident, where left frontal electrodes such as Fp1 and F3 preferentially project toward left posterior sites such as P7, O1, and PO9, whereas right frontal electrodes, including Fp2, F4, and F8, show stronger directed influences toward right parietal and occipital regions, namely P8, O2, and PO10. This lateralized organization is consistent with existing literature linking frontal→parietal alpha connectivity to cognitive control and frontal-occipital alpha interactions to attentional regulation, both of which are known to be modulated under stress conditions.

	\autoref{fig:beta_feature_matrix} summarizes the relative importance of beta-PDC connectivity features, whereas \autoref{fig:beta_directed} illustrates the directed interaction network obtained by retaining only connections with importance values exceeding 0.9. In contrast to alpha-PDC, the beta-PDC connectivity pattern demonstrates prominent involvement of fronto-central, central, and fronto-temporal regions, indicating a prominent contribution from sensorimotor and executive control networks. Electrodes such as Fz, FC1, FC2, C3, C4, and Cz emerge as key sources of directed influence, with further contributions from frontal sites including Fp1 and Fp2, projecting toward parietal and occipital regions such as Pz, O1, O2, Oz, and PO10, which primarily function as receiving nodes. Numerous fronto-central → parietal and central → occipital beta interactions are evident, highlighting the role of beta oscillations in coordinating information flow between motor and sensory regions under stress. Notably, pronounced directed pathways originating from central and fronto-central sites (C3, C4, Cz, FC1, FC2) toward posterior regions (O1, O2, PO10) suggest enhanced stabilization and sustained cognitive engagement, which are characteristic of beta activity during stress-related task performance. A tendency toward lateralized organization is observed, with left central and fronto-central regions preferentially interacting with left posterior sites, and right central regions exhibiting stronger coupling with right parietal–occipital areas. In summary, the beta-PDC topology exhibits greater central involvement than alpha-PDC, reflecting the beta band’s association with motor control, cognitive stability, and sustained attentional states.

	\begin{table}[!htp]
		\centering
		\renewcommand{\arraystretch}{1.2}
		\caption{Comparison between alpha-PDC connections (importance scores $> 0.9$) and those retained under 10\% sparsity threshold.}
		\label{tab:alpha_pdc_sparsity}
		\begin{tabular}{cccc}
			\toprule
			\textbf{Source} & \textbf{Target} & \textbf{Importance Score} & \textbf{Retained} \\
			\midrule
			F8  & FC2 & 1.000000 & Yes \\
			F4  & Pz  & 0.991274 & No \\
			Fp1 & Pz  & 0.986352 & No \\
			Fz  & F3  & 0.968212 & Yes \\
			Fp1 & P7  & 0.964409 & No \\
			Fz  & FC6 & 0.963358 & No \\
			F3  & CP2 & 0.960101 & Yes \\
			F8  & F7  & 0.958841 & Yes \\
			F4  & Cz  & 0.954726 & Yes \\
			F3  & C4  & 0.953343 & Yes \\
			F4  & CP2 & 0.950159 & No \\
			F4  & F7  & 0.947239 & No \\
			F4  & Oz  & 0.943325 & No \\
			F8  & F4  & 0.939283 & Yes \\
			F4  & FC1 & 0.939015 & Yes \\
			F3  & CP1 & 0.923330 & Yes \\
			F7  & FC2 & 0.918216 & Yes \\
			F7  & P7  & 0.914096 & No \\
			F8  & PO9 & 0.913963 & Yes \\
			F7  & F4  & 0.912399 & No \\
			F3  & C3  & 0.907769 & Yes \\
			F8  & F3  & 0.905002 & Yes \\
			Fz  & F7  & 0.904632 & No \\
			Fp2 & F4  & 0.904377 & No \\
			\bottomrule
		\end{tabular}
	\end{table}
	
	To examine whether the discriminative connections identified by the XGB model correspond to physiologically pronounced pathways, sparsity thresholding was applied to the mean alpha-PDC connectivity matrix \citep{bullmore2009complex}. The full directed connectivity matrix of 992 connections was first ranked by mean alpha-PDC strength, and the strongest 10\% of connections were retained to form the sparsity thresholded network. The set of discriminative connections identified by XGB with an importance score above 0.9 was then compared against this sparsity-thresholded network, as shown in \autoref{tab:alpha_pdc_sparsity}. Among the 24 discriminative alpha-PDC connections, 13 were retained within the top 10\% of the strongest directed interactions. Representative overlapping connections include F8 → FC2, Fz → F3, F4 → Cz, and F3 → CP1, indicating prominent frontal-to-central and frontal-to-parietal information-flow pathways in the alpha connectivity. This result indicates that XGB-identified connections also correspond to physiological connectivity pathways in the alpha band, whereas the remaining connections represent weaker yet discriminative interactions that contribute to class separation.
	
	Consequently, the thresholding analysis was performed for the beta-PDC connectivity matrix to assess the relationship between discriminative and physiological connections \citep{bullmore2009complex}. The discriminative connections identified by XGB with an importance score above 0.9 were subsequently compared with this thresholded set. Out of the 21 discriminative beta-PDC connections, 11 were retained within the top 10\% strongest directed interactions as shown in \autoref{tab:beta_pdc_sparsity}. Overlapping connections include F3 → Fz, F4 → FC5, F3 → FC1, and Fp1 → Fp2, reflecting prominent frontal and fronto-central interactions in the beta connectivity. This overlap between the discriminative and sparsity-thresholded networks indicates that several XGB–identified connections correspond to strong physiological information-flow pathways, while others capture subtler yet classification-relevant connectivity variations.
	
	\begin{table}[!htp]
		\centering
		\renewcommand{\arraystretch}{1.2}
		\caption{Comparison between beta-PDC connections (importance scores $> 0.9$) and those retained under a 10\% sparsity threshold.}
		\label{tab:beta_pdc_sparsity}
		\begin{tabular}{cccc}
			\toprule
			\textbf{Source} & \textbf{Target} & \textbf{Importance Score} & \textbf{Retained} \\
			\midrule
			F3  & Fz   & 1.000000 & Yes \\
			F4  & FC5  & 0.995724 & Yes \\
			F3  & FC1  & 0.993127 & Yes \\
			F7  & O1   & 0.988953 & No \\
			Fz  & T7   & 0.986268 & Yes \\
			Fp1 & FC1  & 0.972190 & No \\
			Fp2 & FT10 & 0.970956 & Yes \\
			F7  & FC5  & 0.962776 & No \\
			Fp1 & T7   & 0.960586 & No \\
			F4  & O2   & 0.958211 & No \\
			Fp1 & Fp2  & 0.955383 & Yes \\
			F3  & CP2  & 0.949512 & No \\
			F8  & Oz   & 0.931502 & Yes \\
			F7  & T8   & 0.924786 & Yes \\
			Fp2 & T8   & 0.923433 & No \\
			F7  & Oz   & 0.922622 & No \\
			F3  & F7   & 0.921348 & Yes \\
			F8  & FT10 & 0.920711 & No \\
			F3  & F4   & 0.920005 & Yes \\
			Fp2 & PO10 & 0.914763 & Yes \\
			F8  & P3   & 0.903214 & No \\
			\bottomrule
		\end{tabular}
	\end{table}
	
	The observed frontal-driven alpha-PDC connectivity pattern suggests regulatory mechanisms within cortical networks, whereas fronto-central and central-to-posterior beta- PDC interactions indicate sustained task engagement and sensorimotor stabilization under stress. The consistent emergence of these band-specific patterns across classifiers and modeling strategies further underscores the robustness of alpha and beta-connectivity as reliable neurophysiological markers of stress.
	
	\section{Impact of temporal windows}
	
\begin{figure*}[t]
		\centering
		\includegraphics[width=1.05\textwidth]{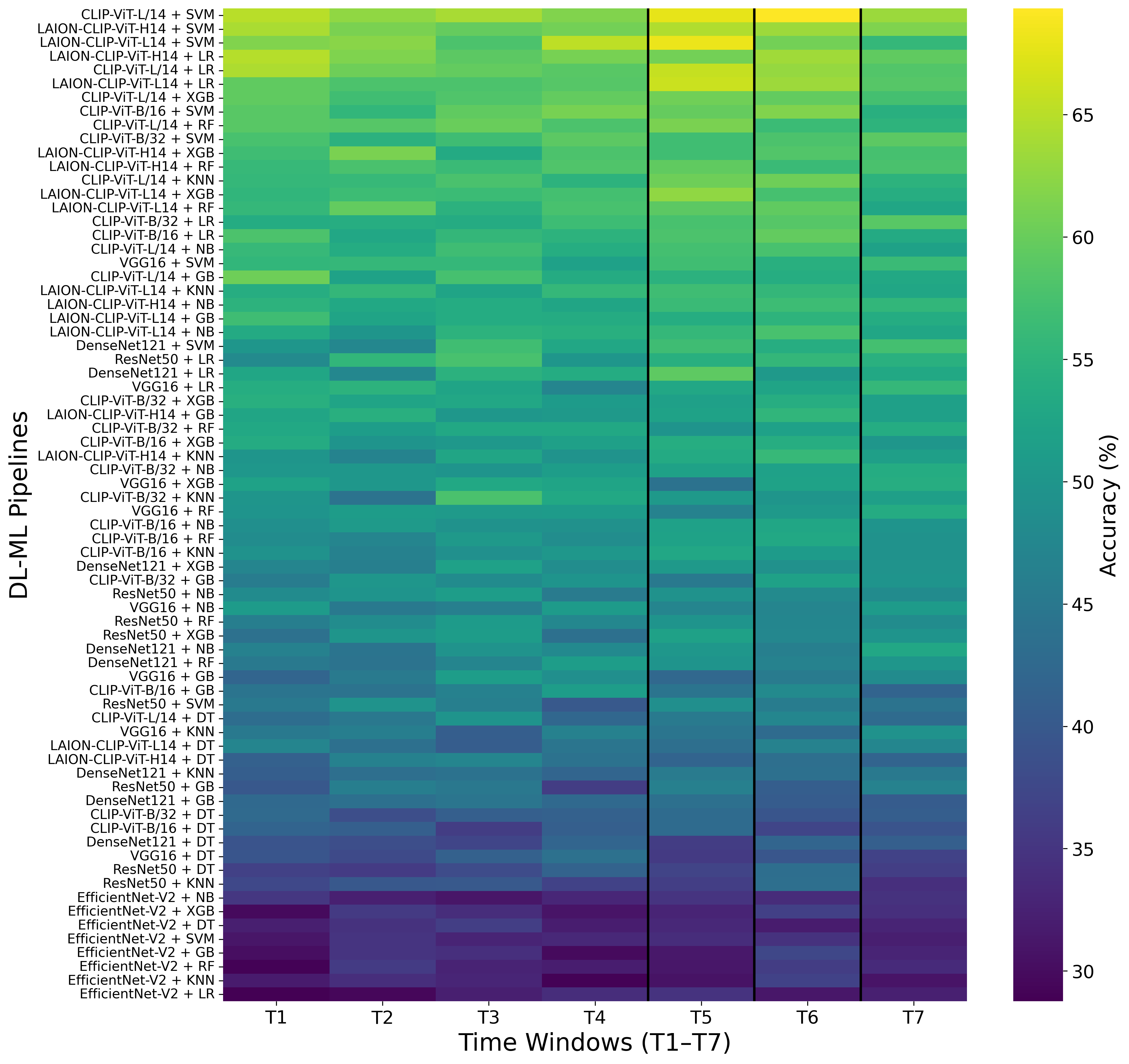}
		\caption{Evolution of temporal windows across DL–ML pipelines.}
		\label{fig:temporal}
	\end{figure*}

	Distinct temporal trends in classification performance are observed across the seven temporal segments T1 to T7, evaluated for all DL-ML combinations. \autoref{fig:temporal} presents the window-wise classification performance of all model combinations. An orderly increase in accuracy is observed from early to central temporal segments, with peak performance consistently occurring in the central-to-late windows T4 to T6 across ViTs and CNN backbones. This temporal progression indicates that stress-related brain activity does not emerge immediately but gradually develops and stabilizes over time. This results in better discrimination between stress states as cognitive demand and autonomic involvement increase during task performance.
	
	Lower performance in the initial windows, such as T1 and T2, likely reflects transient neural adaptation and task familiarization, during which connectivity patterns have not yet fully differentiated across stress levels. As the task progresses, connectivity-based stress markers, such as fronto-parietal alpha and fronto-central beta interactions, become more pronounced and temporally coherent, thereby improving classification performance. The convergence of peak classification accuracies around T5 to T6 across diverse DL-ML pipelines suggests that these windows correspond to a stress state, during which directed connectivity patterns and their learned representations exhibit maximal discriminative power. The global comparative analysis of window-wise performance across all DL-ML pipelines highlights consistent temporal trends, suggesting that intrinsic temporal properties of stress-related brain dynamics drive the observed performance gains. This consistency further motivates a closer examination of temporal performance trends, which clarify how classification accuracy progressively evolves and stabilizes over time within high-performing DL-ML pipelines.
	
	\begin{figure*}[t]
		\centering
		
		\begin{subfigure}{0.49\textwidth}
			\centering
			\includegraphics[width=\linewidth]{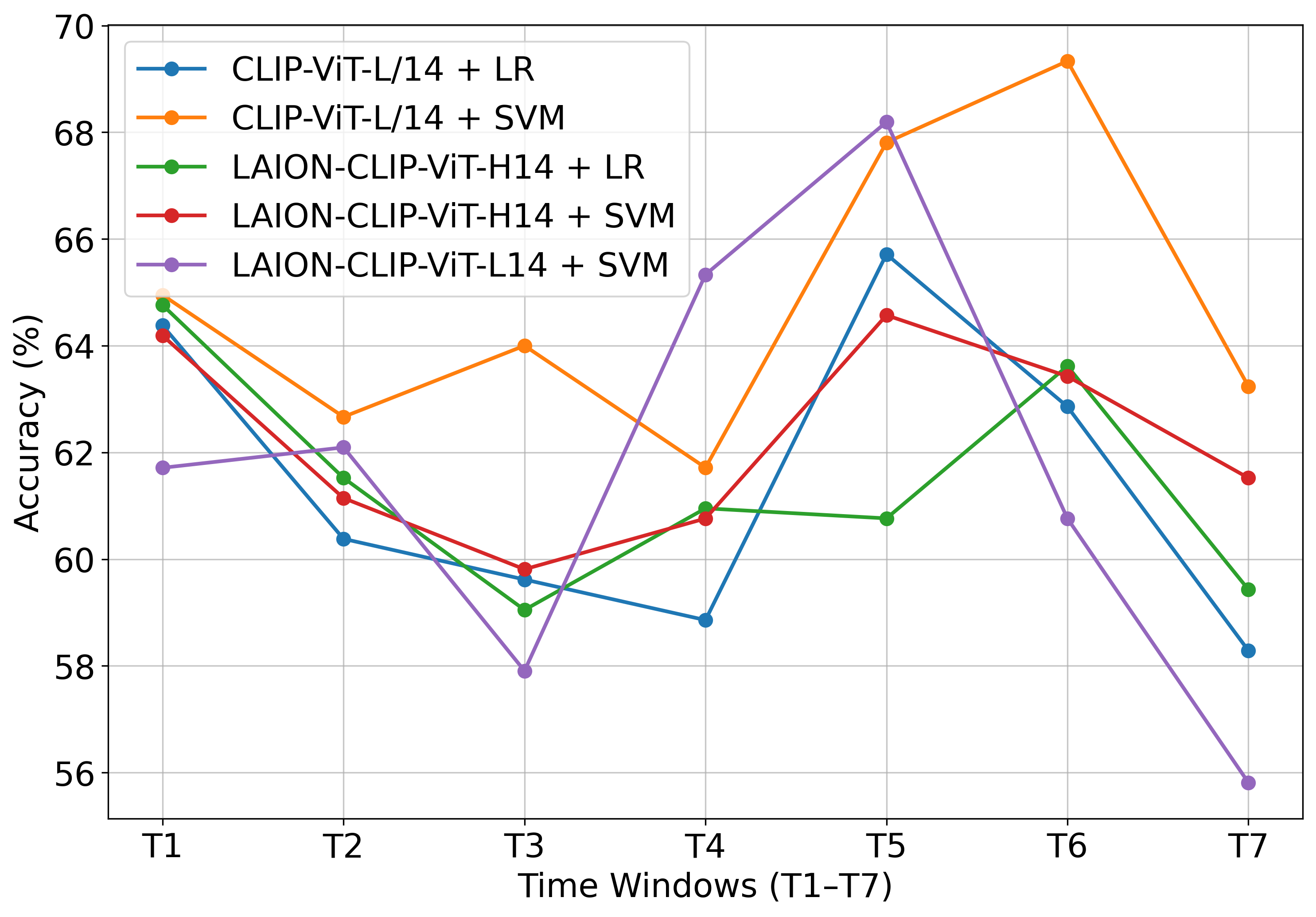}
			\caption{DL–ML pipelines group 1.}
			\label{fig:traj1}
		\end{subfigure}
		\hfill
		\begin{subfigure}{0.49\textwidth}
			\centering
			\includegraphics[width=\linewidth]{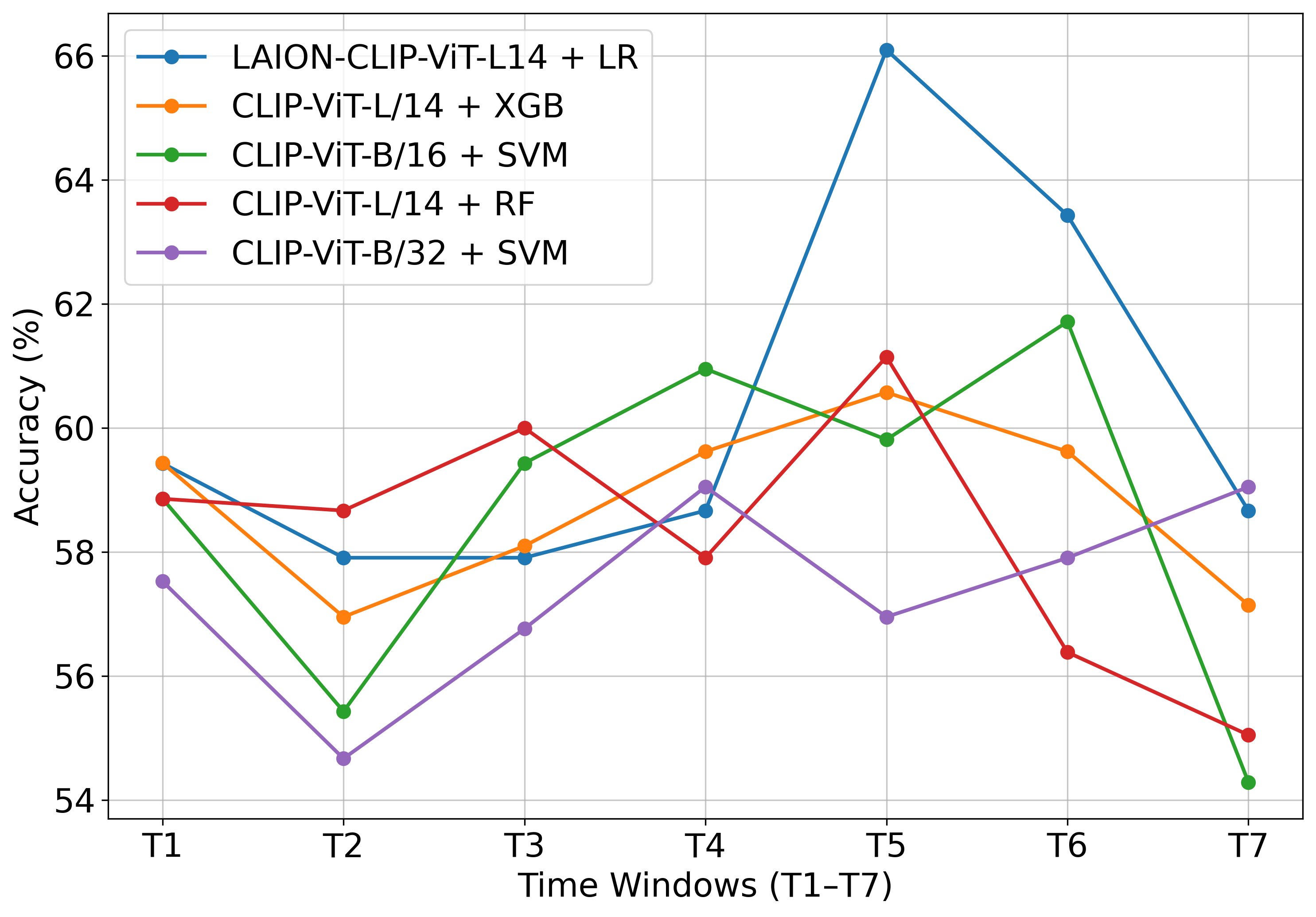}
			\caption{DL–ML pipelines group 2.}
			\label{fig:traj2}
		\end{subfigure}
		
		\vspace{4mm}
		
		\begin{subfigure}{0.49\textwidth}
			\centering
			\includegraphics[width=\linewidth]{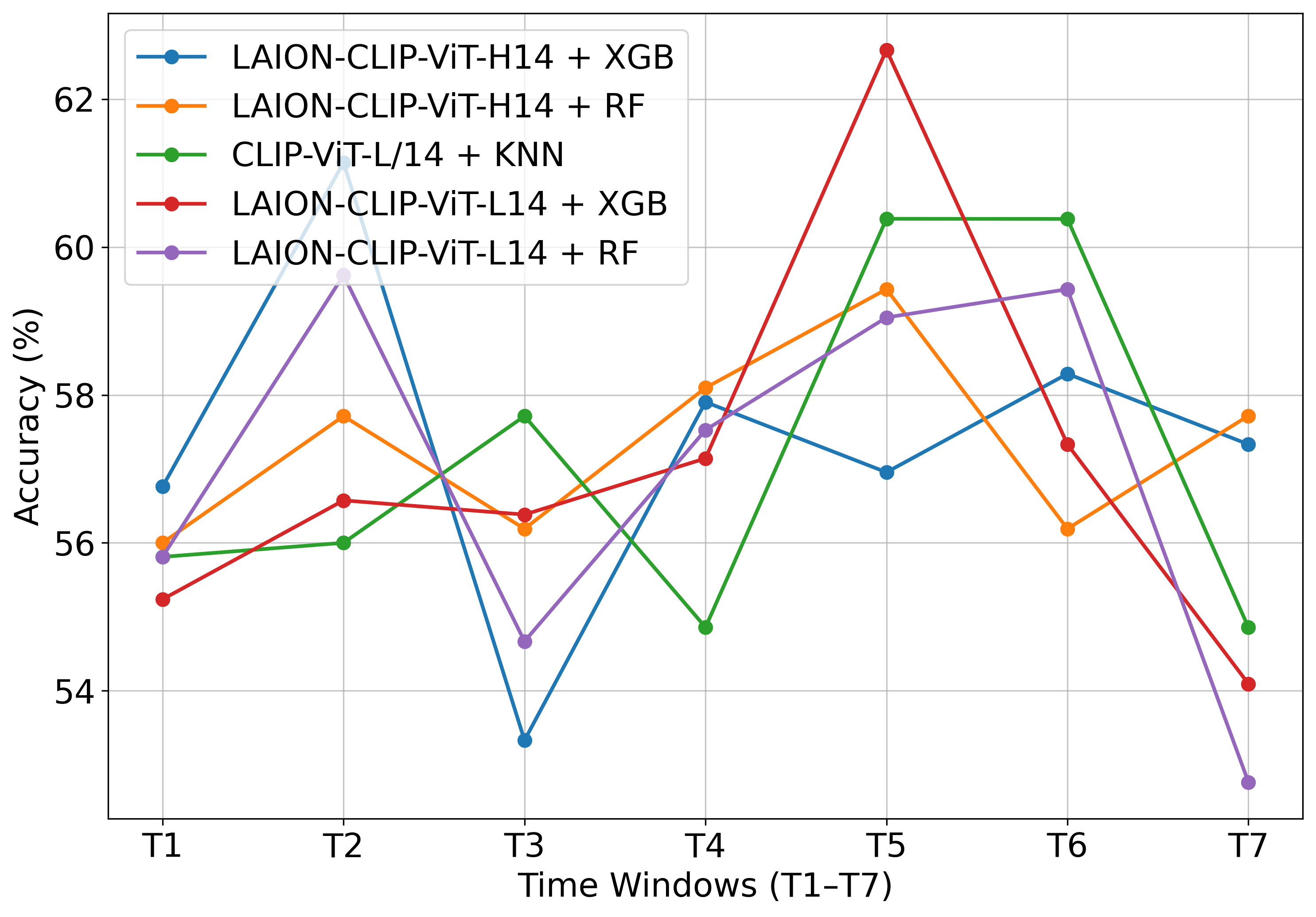}
			\caption{DL–ML pipelines group 3.}
			\label{fig:traj3}
		\end{subfigure}
		\hfill
		\begin{subfigure}{0.49\textwidth}
			\centering
			\includegraphics[width=\linewidth]{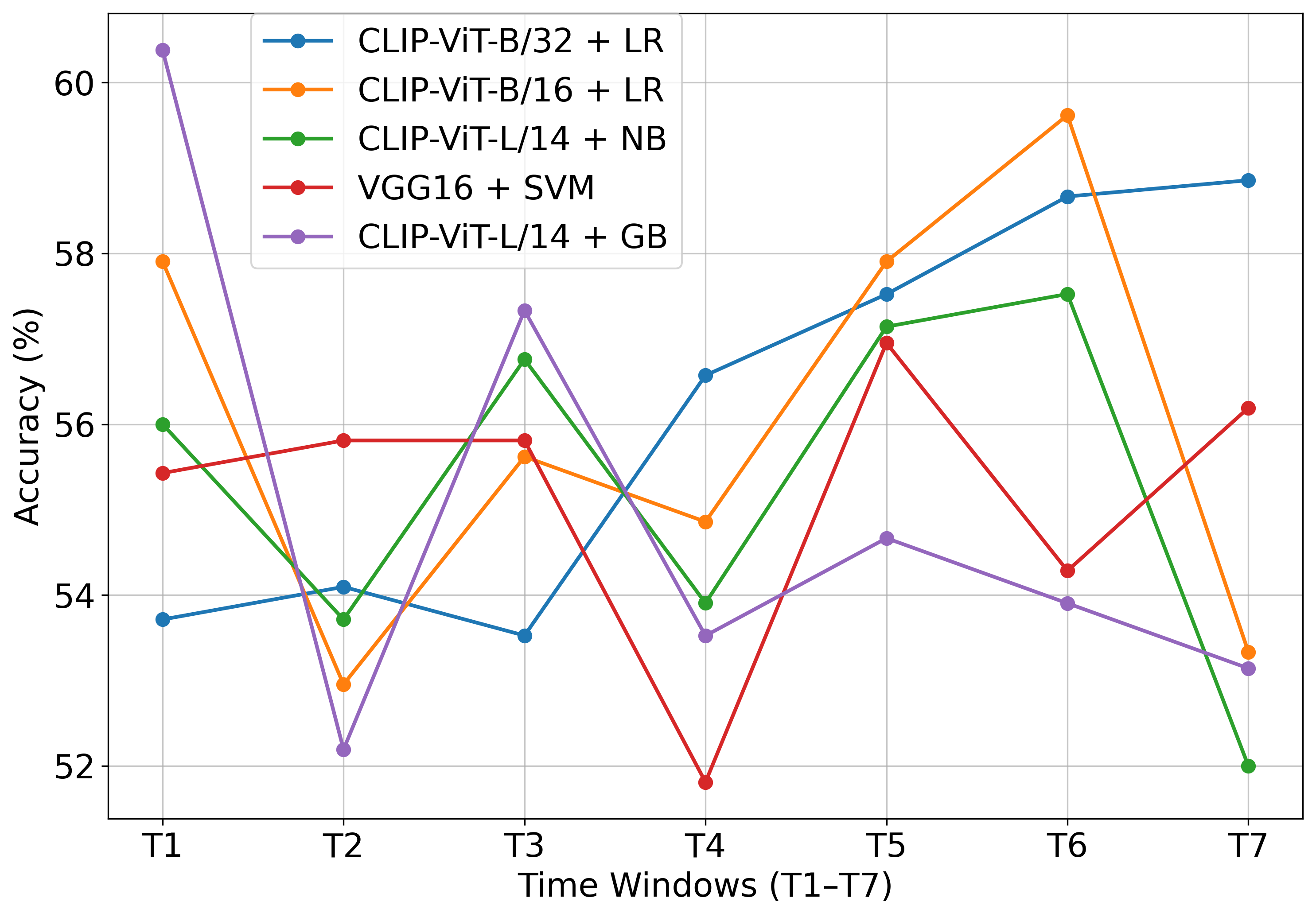}
			\caption{DL–ML pipelines group 4.}
			\label{fig:traj4}
		\end{subfigure}
		
		\vspace{4mm}
		
		\begin{subfigure}{0.49\textwidth}
			\centering
			\includegraphics[width=\linewidth]{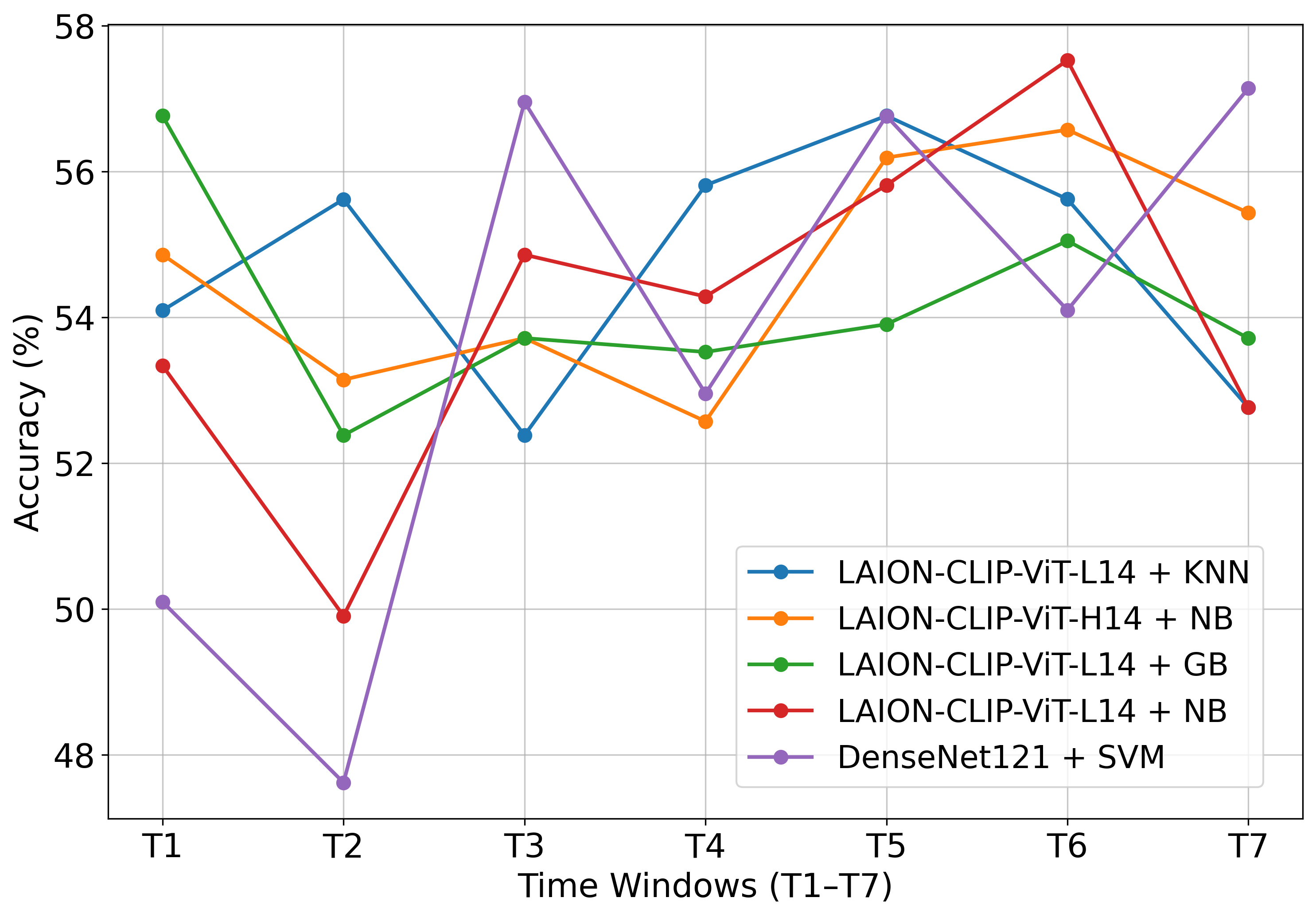}
			\caption{DL–ML group pipelines 5.}
			\label{fig:traj5}
		\end{subfigure}
		
		\caption{Evolution of temporal connectivity patterns among top 20 DL–ML pipelines.}
		\label{fig:trajectories}
		
	\end{figure*}
	
	Accordingly, the top 20 high-performing DL-ML pipelines were selected from the global window-wise comparative analysis in \autoref{fig:temporal} and stratified into 5 convergence groups for detailed investigation, as shown in \autoref{fig:trajectories}. These pipelines were selected based on their overall classification performance across temporal windows, which allows meaningful analysis of temporal trends while maintaining methodological clarity.
	
	Across the 5 hybrid DL–ML groups, as shown in \autoref{fig:traj1} to \autoref{fig:traj5}, a structured evolution in classification patterns is observed, with accuracy generally improving from early windows T1 to T3 towards subsequent windows, particularly T4 to T6. This is followed by mild stabilization or decline at T7. Despite variations in backbone architectures and classifiers, most DL-ML pipelines demonstrate uniform performance peaks within T4 to T6, underscoring the robustness of the temporal segments. The mild or attenuated performance observed at T7 indicates that extended temporal aggregation does not confer additional discriminative benefit. These observations suggest that stress-related connectivity signatures become optimally informative only after sufficient temporal consolidation, independent of specific DL-ML configurations. The connectivity embeddings require a finite temporal span to emerge and stabilize, thereby enabling more reliable learning by downstream classifiers. Consequently, these findings highlight the importance of careful temporal window size selection for optimized classification in dynamic stress-based EEG connectivity frameworks.
	
	\section{Discussion}
	
	Prior studies on stress have reported diverse findings regarding the correlation between stress and EEG frequency bands, including delta, theta, alpha, beta, and gamma oscillations \citep{giannakakis2019review}. Some investigations have documented elevated delta activity under stress, whereas others have observed reductions in delta power during stress exposure \citep{giannakakis2015detection, bosl2023biomarker, chang2023eeg}. Similarly, the theta, alpha, and beta bands have demonstrated inconsistent patterns, with studies reporting both increases and decreases in activity under stress \citep{alonso2015stress, al2015mental, al2017assessment, giannakakis2015detection, bosl2023biomarker, chang2023eeg, minguillon2016stress, acharya2025neurostressology}. Although heightened gamma activity has been reported in certain stress paradigms \citep{minguillon2016stress}, such findings should be interpreted cautiously due to the potential myogenic artifacts \citep{nasteski2017overview, giannakakis2019review}. However, previous studies using the MAT-based SAM 40 dataset have demonstrated the statistical significance of EEG-based PDC features in stress analysis \citep{acharya2025neural}. In particular, beta-PDC exhibited positive correlations with stress levels, whereas alpha-PDC showed statistically significant but negative correlations with connectivity patterns. The main objective of this study is to develop NeuroStrata, a layered analytical pipeline that integrates dynamic effective connectivity estimation with deep neural representations and lightweight ML classifiers to enable hybrid DL-ML modeling of mental stress dynamics.
	
	To explicate the discriminative relevance of TV-PDC connectivity features, XGB was utilized to rank the importance of TV-PDC feature connections, thereby identifying the most influential directed connections contributing to stress classification. From  \autoref{fig:alpha_feature_matrix} and \autoref{fig:alpha_directed}, the feature importance analysis revealed that alpha-PDC was primarily characterized by frontal-driven influences projecting toward parietal and occipital regions, with prominent long-range frontal→posterior interactions extending to central, parietal, and occipital regions, indicating a consistent anterior–posterior directional organization within the stress-related network. These directed alpha pathways suggest the engagement of regulatory control processes governing posterior sensory regions during stress. In contrast, as shown in \autoref{fig:beta_feature_matrix}, beta-PDC exhibited a comparatively integrative, spatially distributed network structure, characterized by strong frontal → central → parietal interactions and pronounced involvement of the central hub. This reflects enhanced cognitive engagement, vigilance, and involvement of sensorimotor-related networks under stress. The distributed beta connectivity further indicates heightened information integration and dynamic network patterns during stress processing. Collectively, these findings suggest that alpha and beta bands encode complementary yet distinct aspects of stress-related brain dynamics, where alpha connectivity reflects regulatory modulation of posterior cortical regions, while beta connectivity captures integrative control and sensorimotor preparedness. The XGB-based feature ranking enhances the neurophysiological interpretability of the proposed NeuroStrata framework and supports TV-PDC as a robust marker of stress-induced directed brain network reorganization. It is important to emphasize that TV-PDC was estimated at the sensor level; therefore, the identified causal influences represent directed connectivity patterns rather than directed anatomical causality and should be interpreted as network-level directional influences rather than source-localized neural pathways.

	Furthermore, distinct temporal trends in classification performance were observed across the seven temporal segments T1 to T7 for all DL-ML combinations, as illustrated in \autoref{fig:temporal}. The classification accuracy progressively improved from early to central temporal windows, with consistent peak performance occurring within T4 to T6 across both CNN and ViT backbones. This temporal progression suggests that stress-related brain dynamics do not emerge instantaneously but gradually develop and stabilize as task engagement increases, thereby improving the separability of stress states. The relatively lower performance in early windows T1 and T2 likely reflects transient neural adaptation and task familiarization, during which connectivity patterns remain insufficiently differentiated across stress levels. As the task progresses, connectivity-based stress markers, particularly fronto-parietal alpha and fronto-central beta interactions, become more temporally coherent, thereby enhancing discriminative capability. The convergence of peak classification performance around T5 and T6 across diverse DL-ML pipelines indicates that these windows correspond to a temporally consolidated stress-responsive state, where directed connectivity patterns and their learned representations attain maximal separability. Supplementary investigation of the temporal patterns of the top 20 high-performing DL-ML pipelines shown in \autoref{fig:trajectories} revealed a structured evolution in classification trajectories, with performance generally increasing from early windows (T1 to T3) toward central windows (T4 to T6), followed by mild stabilization or slight decline at T7. This mitigation at T7 suggests that extended temporal aggregation does not provide additional discriminative advantage, indicating that connectivity embeddings require an optimal temporal span to emerge and stabilize. These findings underscore the importance of temporal window selection in dynamic EEG connectivity frameworks, indicating that stress-related connectivity signatures become maximally discriminative only after adequate temporal consolidation and are largely independent of specific DL-ML configurations. Although TV-PDC incorporates temporal modeling through the STW approach, the observed improvements cannot be attributed solely to temporal dynamics. Instead, the performance gains likely reflect the combined influence of temporal modeling, directionality, and connectivity formulation.
	
\begin{figure*}[t]
		\centering
		
		\begin{subfigure}{0.49\textwidth}
			\centering
			\includegraphics[width=\linewidth]{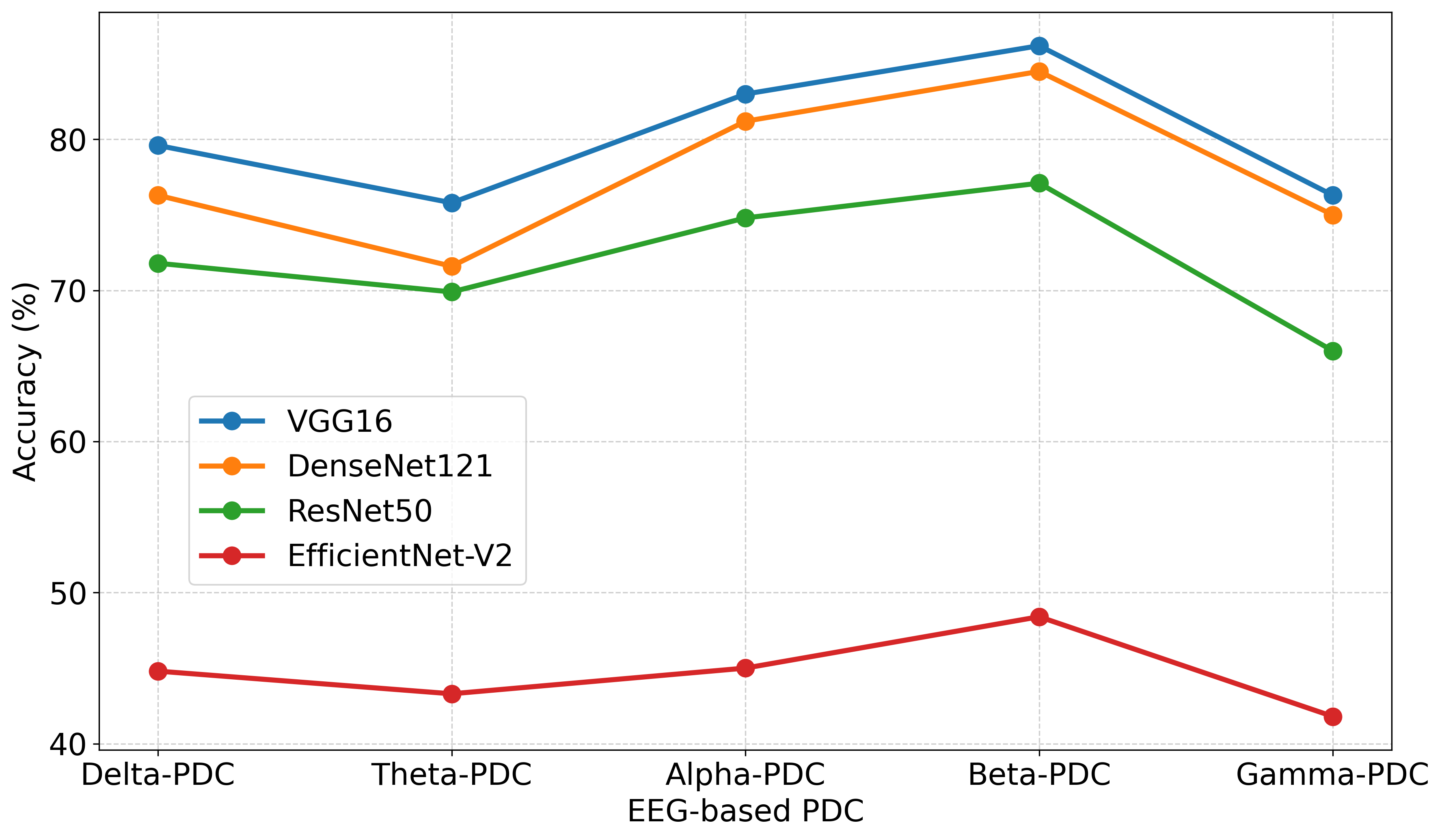}
			\caption{CNN performance trend across EEG-based PDCs.}
			\label{fig:cnn_trend}
		\end{subfigure}
		\hfill
		\begin{subfigure}{0.49\textwidth}
			\centering
			\includegraphics[width=\linewidth]{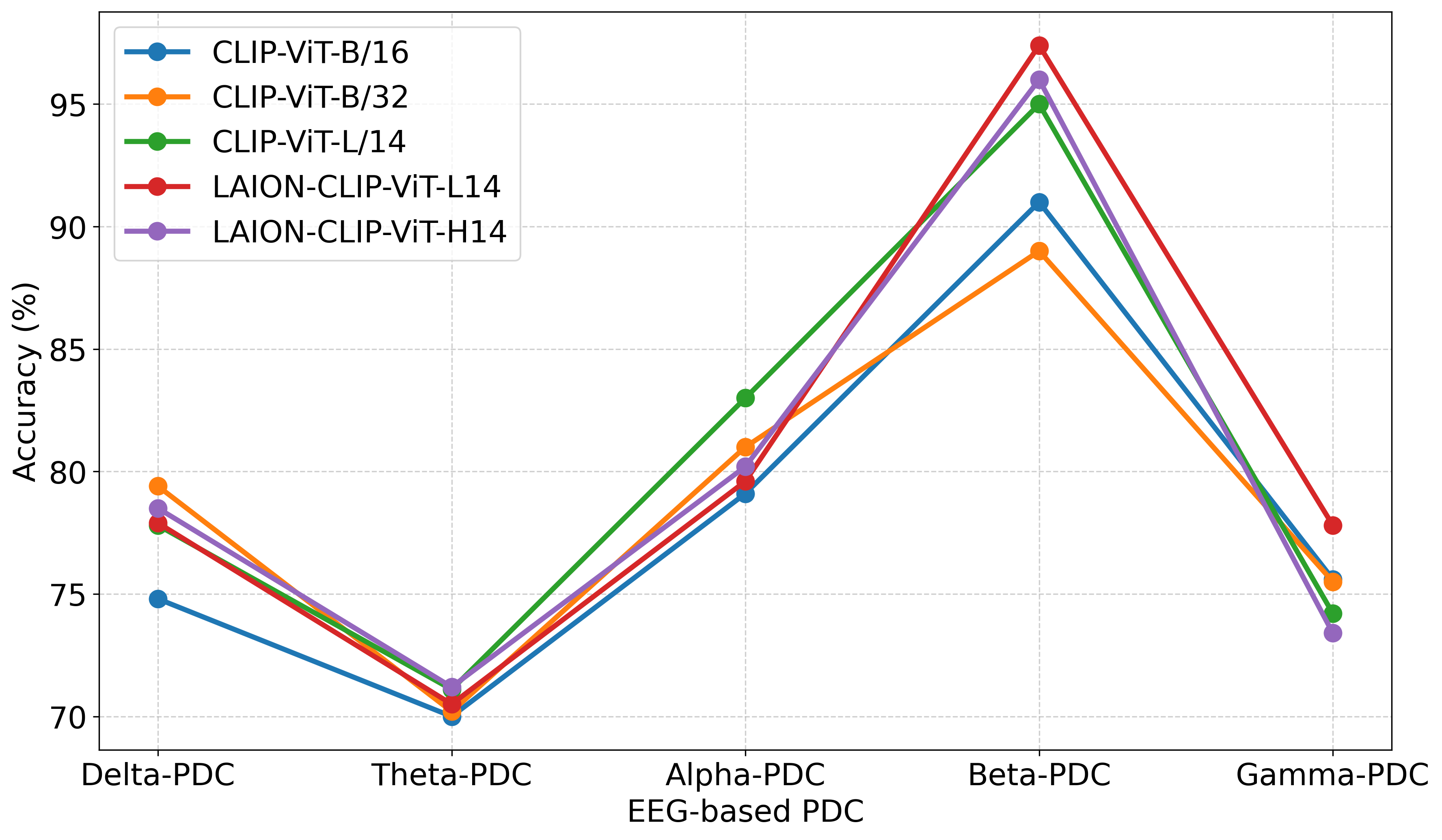}
			\caption{ViT performance trend across EEG-based PDCs.}
			\label{fig:vit_trend}
		\end{subfigure}
		
		\caption{Performance comparison trends of CNN and ViT models across EEG-based PDC features.}
		\label{fig:dl_trends}
		
	\end{figure*}

	The performance trend for 3-class stress classification across ViT models on TV-PDC features is shown in \autoref {fig:vit_trend}. To analyze the intrinsic discriminative strength of each ViT backbone independently of classifier-specific bias, the highest classification accuracy achieved across all evaluated ML classifiers was used for each EEG-frequency-specific PDC feature set. This strategy provides an estimate of the strongest observed representational capability of the learned connectivity embeddings and supports a representation-focused comparison of backbone effectiveness while minimizing dependence on any single classifier. As illustrated by the ViT performance trend, a consistent hierarchical pattern is observed across frequency bands, with larger-capacity models generally yielding stronger discriminative performance. All ViT models exhibit a clear performance peak in the beta-PDC band, confirming that beta-band connectivity encodes the most separable stress-related information. Among the models, LAION-CLIP-ViTL14 achieves the highest peak, followed closely by LAION-CLIP- ViT-H14 and CLIP-ViT-L/14, while the base variants show comparatively lower performance. This order remains relatively stable across alpha and gamma bands, whereas theta-PDC consistently represents the weakest band, indicating lower separability of connectivity patterns. This consistent capacity-dependent behavior demonstrates that larger ViT backbones learn more informative connectivity embeddings, and the stability of the trend across EEG bands supports the robustness of the learned representations.
	
	Similarly, the performance trend for 3-class stress classification across the CNN models using TV-PDC features is shown in \autoref {fig:cnn_trend}. To evaluate the representational strength of the CNN backbones independently of classifier-specific effects, the maximum classification accuracy achieved across all considered ML classifiers was selected for each EEG frequency-specific PDC feature set. This strategy reflects the strongest observed discriminative performance of the extracted connectivity features and enables a representation-focused comparison of backbone effectiveness. By avoiding reliance on any single classifier, this approach reduces classifier-dependent bias and allows the comparison to focus primarily on representation quality rather than classifier optimization effects. The resulting trend indicates consistently strong performance of VGG16 across all EEG-based PDC analyses, with DenseNet121 consistently emerging as the second-best architecture, while ResNet50 demonstrates moderate performance, and EfficientNet-V2 remains comparatively weaker. A pronounced improvement in beta-PDC is observed across all CNN models, highlighting the strong discriminative power of beta-based effective connectivity. The alpha-PDC also demonstrates relatively high separability, whereas performance declines for the delta-PDC and gamma-PDC, reaching its lowest level for the theta-PDC. This pattern indicates limited discriminatory information in the theta band. Unlike the ViT models, the CNN hierarchy appears more uniform across EEG bands, suggesting stable yet comparatively lower representational capacity. These findings indicate that while conventional CNN architectures can effectively capture spatial patterns in connectivity maps, their discriminative capability is consistently surpassed by transformer-based representations, particularly for higher-information connectivity, such as beta-PDC.
	
	\begin{table*}[p]
		\centering
		\caption{Comparative analysis of stress studies based on EEG connectivity features.}
		\label{tab:comp_connectivity}
		
		\normalsize
		\renewcommand{\arraystretch}{1.25}
		\setlength{\tabcolsep}{3.2pt}
		
		\adjustbox{max width=\textwidth}{%
			\begin{tabular}{c c c c c C{2.4cm} C{2cm} C{1.5cm} C{2.1cm} C{3cm}}
				\toprule
				\textbf{Ref.} &
				\textbf{Dataset used} &
				\textbf{Experimental Stress tasks} &
				\textbf{No. of EEG Channels} &
				\textbf{No. of Subjects} &
				\textbf{Connectivity Features} &
				\textbf{Key electrodes} &
				\textbf{Top Classifier (Class)} &
				\textbf{Best Accuracy Achieved (Feature) (Class)} &
				\textbf{Observations} \\
				\midrule
				
				\citep{alonso2015stress} &
				Custom &
				MAT &
				19 &
				30 &
				Coherence, MSC, CMIF &
				Not identified &
				Not classified &
				Not classified &
				Reports scalp connectivity maps from coherence and CMIF analyses, showing increased high-beta coherence and decreased alpha coherence in anterior cortical regions. \\[1mm]
				
				\citep{khosrowabadi2018stress} &
				Custom &
				Long-term examination &
				8 &
				26 &
				DTF, dGC, PSI &
				C3, C4, F4, P3, P4, T8 &
				SVM (4) &
				90.90\% (PSI) (4) &
				PSI analysis identified prominent hub regions, with C4 during positive emotional states and P3 during negative emotional states. \\[1mm]
				
				\citep{balconi2018functional} &
				Custom &
				Competition / social stress &
				15 &
				14 &
				Correlation-based linear brain connectivity, MSEC, GC, PSI &
				Not identified &
				Not classified &
				Not classified &
				Partial correlation analysis yields undirected measures of functional connectivity. \\[1mm]
				
				\citep{darzi2022brain} &
				Custom &
				LAPS + audio clips &
				8 &
				26 &
				GC, DTF &
				P3 and T8 &
				SVM (2) &
				Above 90\% (GC) (2); Above 90\% (DTF) (2) &
				The P3$\rightarrow$T8 pathway connection was identified as the most prominent and discriminative connectivity feature. \\[1mm]
				
				\citep{hag2021eeg} &
				Custom &
				MAT &
				7 &
				22 &
				Delta-PLV, Theta-PLV, Alpha-PLV, Sigma-PLV, Low beta-PLV, High beta-PLV &
				P7, F3, Fz, F8, Fp1, Fp4 &
				LDA (2) &
				75.20\% (Delta-PLV) (2); 71.90\% (Alpha-PLV) (2); 73.40\% (High beta-PLV) (2) &
				PLV is an undirected connectivity measure; although significant frontal channel involvement was reported, directional information flow was not captured.\\[1mm]
				
				\citep{al2021prefrontal} &
				Custom &
				MAT &
				7 &
				25 &
				Alpha-based Coherence &
				F3, F4, F7, F8, Fp1, Fp2 &
				Not classified &
				Not classified &
				Reports a pronounced dominance of the right dorsolateral prefrontal cortex in the alpha band. \\[1mm]
				
				\citep{vanhollebeke2023effects} &
				Custom &
				Visual reasoning puzzles &
				57 &
				73 &
				Alpha-based power, Amplitude Envelope Correlation (AEC) &
				Not identified &
				Not classified &
				Not classified &
				Reports significant increase in alpha power in the left precuneus, right precuneus, and right posterior cingulate cortex. \\[1mm]
				
				\textbf{Proposed Study} &
				SAM 40 &
				MAT &
				32 &
				35 &
				Delta-PDC, Theta-PDC, Alpha-PDC, Beta-PDC, Gamma-PDC &
				Alpha: frontal (Fp1, Fp2, F3, F4, F7, F8), central (C3, C4, Cz), posterior (P3, P4, O1, O2);
				Beta: frontal--central integration &
				LAION-CLIP-ViT-L14 + SVM (3) &
				97.3\% (Beta-PDC) (3) &
				Reports alpha connectivity with long-range $frontal \rightarrow parietal$, 
				$frontal \rightarrow central$, and $frontal \rightarrow occipital$ connections. 
				Beta connectivity with $frontal \rightarrow frontal$, 
				$frontal \rightarrow parietal$, and $frontal \rightarrow occipital$ connections. \\
				
				\bottomrule
			\end{tabular}%
		}
	\end{table*}
	
	The Dataset Used column in \autoref{tab:comp_connectivity} distinguishes between openly available datasets, such as the SAM 40 dataset \citep{ghosh2022sam}, and the custom EEG datasets collected for individual studies. Although these datasets are not publicly distributed, the EEG signals were acquired using standardized and reproducible experimental paradigms, including arithmetic tasks, Stroop paradigms, and audio–visual stimulation protocols. Accordingly, equivalent datasets can be replicated under analogous experimental conditions, thereby partially mitigating constraints arising from limited public data availability.
	
	In the EEG-based stress classification studies summarized in \autoref{tab:comp_connectivity}, prior research indicates that connectivity measures such as Granger Causality (GC) and Directed Transfer Function (DTF) achieved classification accuracies exceeding 90\% for binary stress detection using 8 EEG channels across data from 26 participants \citep{darzi2022brain}. Similarly, the Phase Slope Index (PSI) reported an accuracy of 90.90\% in a four-class stress classification setting under the same channel and participant configuration \citep{khosrowabadi2018stress}. However, these connectivity features do not simultaneously capture directional influences among EEG channels with frequency-specific resolution, thereby limiting their ability to characterize stress-modulated information transfer across brain regions within distinct EEG frequency bands. Such limitations may result in the loss of critical insights into time-varying neural information flow during stress conditions, which is fundamental for EEG-based neurological studies. Furthermore, another study employing 7 EEG channels across 22 participants reported classification accuracies of 75.20\%, 71.90\%, and 73.40\% for binary stress detection using frequency-specific functional connectivity based on Phase Locking Value (PLV) in the delta, alpha, and high-beta bands, respectively  \citep{hag2021eeg}. Nevertheless, PLV quantifies phase synchronization between EEG channels rather than directed interactions. In a similar context, Amplitude Envelope Correlation (AEC), another functional connectivity metric, has been used to capture statistical interdependence among brain regions \citep{vanhollebeke2023effects}; however, it does not capture information regarding frequency-resolved or time-varying directional neural dynamics. In contrast, the proposed NeuroStrata framework, based on Time-Varying Partial Directed Coherence (TV-PDC), captures both frequency-specific characteristics and dynamic directional information flow, achieving improved classification performance of 97.3\% accuracy using the LAION-CLIP-ViT-L14 backbone with SVM classifier on Beta-PDC features. This performance also surpasses all CNN-based configurations, where the best result reached 86.0\% accuracy using the VGG16 backbone with an LR classifier on Beta-PDC features. These findings highlight the superior representation capability of large-scale ViT embeddings over conventional CNN features for modeling dynamic stress-related EEG connectivity.
	
	\subsection{Advantages and limitations}
	
	The proposed NeuroStrata pipeline provides a structurally layered and neurophysiologically grounded framework that integrates time-varying, frequency-specific, and directed causal influences with high-performing deep embedding learning. By modeling time-varying PDC, the pipeline captures temporally evolving Granger-causality-based information-flow patterns across brain networks. This characterizes dynamic network reconfiguration rather than static connectivity snapshots. The dual-backbone design, with CNNs for localized spatial pattern extraction and ViTs for modeling long-range dependencies, forms a complementary representation framework that jointly encodes effective connectivity and global network topology within directed connectivity matrices. Dimensionality reduction through fold-wise PCA further reduces redundancy while preserving variance-oriented embedding subspaces, thereby improving statistical stability in high feature-to-sample schemes and preventing cross-validation information leakage. The subsequent lightweight ML layer enables flexible learning of decision boundaries over compact latent connectivity manifolds, allowing nonlinear, probabilistic, and ensemble-based modeling without retraining large deep networks. Collectively, this convergence architecture yields highly discriminative yet interpretable embeddings, supports EEG band-specific analysis, preserves the temporal evolution of neural dynamics, and demonstrates strong discriminative capability as evidenced by peak classification performance using PDC-based directed connectivity.
	
	On the contrary, despite the strengths, the pipeline exhibits several methodological constraints. Firstly, TV-PDC is estimated in the sensor space, meaning it infers directional influences of EEG electrodes rather than source-localized neural causation. This study was confined to the SAM 40 dataset comprising 35 young, neurologically healthy participants, which may constrain the broader applicability of the findings to clinical or medical populations \citep{ghosh2022sam}. Although the alpha-PDC and beta-PDC demonstrated strong discriminative capability, comparatively lower performance in the delta, theta, and gamma bands indicates frequency-dependent variability in stress-related EEG signatures. The framework further relies on fixed temporal segmentation parameters, specifically a 5-second window with a 2-second overlap, which may not optimally capture transient neural dynamics under alternative stress paradigms. Reliance on pretrained backbones such as CNNs and ViTs may introduce a domain-distribution mismatch, as the models pretrained on natural image statistics are not explicitly optimized for structured connectivity matrices; although transferable representations often remain effective, the absence of task-specific adaptation may limit optimal encoding of connectivity topology. In addition, inter-subject variability was not explicitly modeled, despite its potential influence on classification robustness across heterogeneous populations. The exclusive reliance on connectivity features, without incorporating complementary physiological modalities such as EDA or ECG, also limits the results' multi-modal interpretability. Lastly, SAM 40 provides EEG data recorded in a controlled offline environment; therefore, additional investigation is required to extend the NeuroStrata framework toward real-time stress monitoring under diverse stress-inducing conditions.
	
	\section{Conclusion}
	
	This research introduces NeuroStrata, an EEG connectivity driven deep representation learning framework for stress quantification using dynamic effective connectivity derived from Time-Varying Partial Directed Coherence (TV-PDC). The results demonstrate that incorporating directional and frequency-specific temporal connectivity significantly improves stress discrimination compared with conventional static connectivity approaches. Among the EEG bands, beta-PDC exhibited the strongest performance, achieving an accuracy of 97.3\% using the LAION-CLIP-ViT-L14 feature extractor combined with an SVM classifier, while alpha-PDC consistently provided stable and reliable discrimination across different DL–ML model configurations. Furthermore, XGB-based feature importance analysis revealed prominent long-range frontal-to-parietal and frontal-to-occipital alpha interactions, along with frontal–central–parietal beta integrations, highlighting structured anterior–posterior regulation and enhanced inter-regional coordination as key connectivity signatures underlying stress-related brain dynamics. Temporal analysis further indicates that stress-sensitive connectivity patterns become maximally informative only after sufficient temporal consolidation, emphasizing the importance of appropriate temporal window selection for effective stress classification. To conclude, the NeuroStrata framework provides reliable, interpretable, and highly informative features for EEG-based stress analysis, while advancing the neurophysiological understanding of stress and supporting the potential of EEG-driven mental health monitoring systems.
	
	\section*{CRediT authorship contribution statement}
	
	\noindent
	\textbf{Sayantan Acharya:} Writing – original draft, Conceptualization, Visualization, Validation, Software, Resources, Methodology. 
	\textbf{Hamzeh Asgharnezhad:} Visualization, Software, Methodology.
	\textbf{Abbas Khosravi:} Writing – original draft, Conceptualization, Visualization, Supervision, Formal analysis, Data curation. 
	\textbf{Douglas Creighton:} Writing – original draft, Project administration, Formal analysis. 
	\textbf{Roohallah Alizadehsani:} Writing – original draft, Supervision, Resources, Conceptualization. 
	\textbf{U. Rajendra Acharya:} Writing – original draft, Supervision, Project administration, Methodology, Investigation.
	
	\section*{Funding}
	\noindent
	No funding was received for this study.
	
	\section*{Declaration of Competing Interest}
	\noindent
	The authors declare that they have no known competing financial interests or personal relationships that could have appeared to influence the work reported in this paper.
	
	\bibliographystyle{cas-model2-names}
	\bibliography{References}
	
\end{document}